\documentclass[11pt, twocol]{article}

\usepackage{balance}       % to better equalize the last page
\usepackage{graphics}      % for EPS, load graphicx instead
\usepackage[T1]{fontenc}   % for umlauts and other diaeresis
\usepackage{txfonts}
\usepackage{mathptmx}
\usepackage[pdflang={en-US},pdftex]{hyperref}
\usepackage{color}
\usepackage{booktabs}
\usepackage{textcomp}

\usepackage{microtype}        % Improved Tracking and Kerning
\usepackage{ccicons}          % Cite your images correctly!
\usepackage{todonotes}

\def\plaintitle{CardKit: A Card-Based Programming Framework for Drones}

\def\emptyauthor{}
\def\plainkeywords{Drones; Internet of Things; Card-based Programming.}

\makeatletter
\def\url@leostyle{%
  \@ifundefined{selectfont}{
    \def\UrlFont{\sf}
  }{
    \def\UrlFont{\small\bf\ttfamily}
  }}
\makeatother
\def\pprw{8.5in}
\def\pprh{11in}

\definecolor{linkColor}{RGB}{6,125,233}
\hypersetup{%
  pdftitle={\plaintitle},
  pdfauthor={\emptyauthor},
  pdfkeywords={\plainkeywords},
  pdfdisplaydoctitle=true, % For Accessibility
  bookmarksnumbered,
  pdfstartview={FitH},
  colorlinks,
  citecolor=black,
  filecolor=black,
  linkcolor=black,
  urlcolor=linkColor,
  breaklinks=true,
  hypertexnames=false
}

\usepackage{soul}
\usepackage{pbox}
\usepackage{tabularx}
\begin{document}

\title{\plaintitle}

%\numberofauthors{3}
%\author{%
%  \alignauthor{Saad Ismail\\
%    \affaddr{IBM Research~\footnote{This project was completed while an intern and full-time employee of IBM Research.}}\\
%    \affaddr{Yorktown Heights, NY 10598}\\
%    \email{saad.ismailm@gmail.com}}\\
%  \alignauthor{Justin G. Manweiler\\
%    \affaddr{IBM Blockchain}\\
%    \affaddr{Yorktown Heights, NY 10598}\\
%    \email{jmanweiler@us.ibm.com}}\\
%  \alignauthor{Justin D. Weisz\\
%    \affaddr{IBM Research AI}\\
%    \affaddr{Yorktown Heights, NY 10598}\\
%    \email{jweisz@us.ibm.com}}\\
%}

\author{Saad Ismail~\footnote{Author was an intern with IBM Research when this work was completed.}, Justin G. Manweiler, Justin D. Weisz \\
IBM Research \\
Yorktown Heights, NY 10598 \\
\small{saad.ismailm@gmail.com, jmanweiler@us.ibm.com, jweisz@us.ibm.com}
}
\date{}

\maketitle

\begin{abstract}
Drones are being used in many industries for a variety of applications, including inspecting bridges, surveying farm land, and delivering cargo. Automating these kinds of scenarios requires more than following a sequence of GPS waypoints; they require integrating on-device hardware with real-time analysis to provide feedback and control to the drone. Currently, implementing these kinds of advanced scenarios is a complex task, requiring skilled software engineers programming with drone APIs. We envision an alternate model to enable drone operators to orchestrate advanced behaviors using a card-based approach. We describe the design of our card-based programming model, position it relative to other visual programming metaphors, share results from our paper prototype user study, and discuss our learnings from its implementation. Results suggest that a wide range of scenarios can be implemented with moderate mental effort and learning, balanced by intuitiveness and engagement.
\end{abstract}

%\category{H.5.m.}{Information Interfaces and Presentation
%  (e.g. HCI)}{Miscellaneous}

%\keywords{\plainkeywords}

\section{Introduction}
\label{sec:intro}

\begin{figure}[tb]
\centering
\includegraphics[width=2.4in]{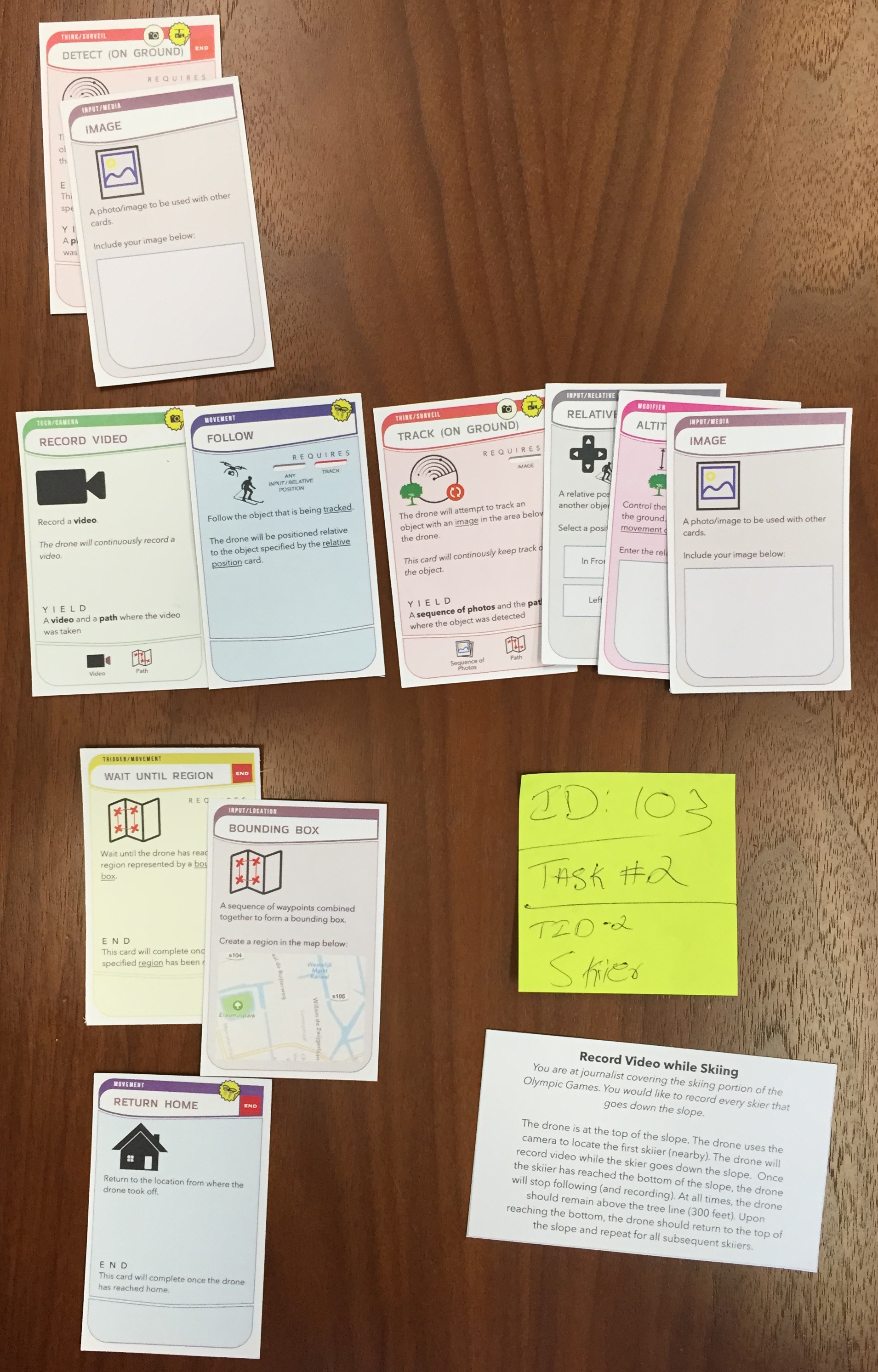}
\caption{Card program instructs a drone to follow and record video of a skiier. This program is constructed as a sequence of four \emph{hands}, sequenced vertically.}
\label{fig:implemented_scenario}
\vspace{-1em}
\end{figure}

Unmanned Autonomous Vehicles (UAVs) or ``drones'' have begun to radically transform many industries. Drones are already being used for applications such as delivering packages, recording feature films, inspecting cell towers and wind turbines, mapping construction sites, and even shepherding and lifeguarding. Drones are especially advantageous for tasks that are dangerous (e.g., bridge inspection, forest fire mapping) or expensive (e.g., aerial surveying).

Because of the huge potential of drones to perform work tasks with increased safety and efficiency, we envision a future of diverse and complex drone applications. Some of these applications are already starting to emerge, such as using drones for pizza, medicine, and package delivery, recording feature films, inspecting cell towers and wind turbines, and even shepherding and lifeguarding. These applications are dynamic and heterogeneous, and are typically ``implemented'' by having one or more human operators control the drone to perform the task. Mobile apps can ease the burden of controlling a drone's flight by pre-programming it to fly on specified paths, or to perform specific camera maneuvers (e.g. follow me, cable cam), but these apps tend to only work well for a single use case and do not adapt to alternative use cases.

At the heart of these applications is a worker who wishes to use a drone as a tool to automate a portion of their job. Many drones have programming APIs, and many of the applications mentioned above could be implemented with these APIs. However, drone APIs are typically complex and require knowledge of coding to use, making them inaccessible to those workers who do not have the requisite skills. For example, a farmer may wish to use their drone to find lost cattle, but building that application requires combining knowledge of a drone API and computer vision algorithms with the skill to implement it across a drone and a mobile device. Therefore, we perceive both a need and an opportunity to design a simple and intuitive ``programming'' paradigm for drones that can automate a diverse set of drone behaviors.

As human-computer interaction researchers, our instincts led us to explore simplified and pedagogical programming environments, such as those aimed toward young or inexperienced programmers~\cite{Bau:2017, Malan:2007, Carlisle:2005}. However, these languages typically expose low-level programming constructs such as conditional logic, looping and recursion, and data structures. Although providing Turing completeness, such sources of complexity \emph{may not actually} be required to implement a number of complex drone scenarios. Instead, our aim is to create a programming paradigm in which \emph{ease of use} trumps \emph{computability} when such a tradeoff must be made.

%We believe many complex behaviors enabled by low-level programming constructs can also be achieved through a higher-level abstraction of a ``card.'' A card is designed to perform a high-level action or behavior such as data gathering, data processing, logical control-flow operations, or hardware manipulations (e.g. flying to a location, taking a photo). Our notion of a card is spiritually similar to other ``cards'' that have been used for programming robots~\ref{zhao2009}. It is a higher-level abstraction than the ``blocks'' used in block-based programming languages~\ref{Bau:2017}.
Many of the programmable interactions with robots and other IoT devices concern its actions and behaviors. For example, a device may gather data (sense), process data (think), or manipulate its hardware state (act/communicate). These activities are high-level abstractions that mask the complexity of the underlying task (e.g. performing a computer vision task on an image, or driving a servo motor to achieve a desired movement). Our aim is to keep the \emph{programming} of these activities at a suitably high level. Block-based metaphors~\cite{Bau:2017} can mask some of these complexities, but block-based languages still expose lower-level programming constructs such as variables, loops, and functions. We questioned whether these constructs were truly necessary for writing useful, non-trivial programs. In our ideation, we settled on the notion of a ``card'' to encapsulate a high-level behavior for sensing, thinking, acting, or communicating. Our notion of a card is spiritually similar to other ``cards'' developed for programming robots~\cite{zhao2009}, but it eliminates some lower level programming constructs (variables, loops, functions), and changes the use of others (conditional logic), used in block-based programming languages.

We readily admit that we did not attempt to devise a general-purpose programming language with strong theoretical underpinnings; rather, our aim was to create something easy to use for a wide, but possibly incomplete, range of scenarios. We examined a broad spectrum of programming abstractions, both pedagogical and for practitioners (see related work), and we concentrated on those that we believed would map well to mobile devices -- convenient for ``in the field'' operations. Our insight was that ``cards'' was a good metaphor because many card games enable a player to implement complex strategies within constraining rule systems, while remaining usable on a mobile device.

Our vision for a card-based programming model converged across many iterations of design, paper prototyping, and testing. Our use of the physical medium enabled rapid design exploration, evaluation, and iteration. In this paper we only report on the final design upon which we converged; discussion of the numerous design options we attempted and rejected is beyond our scope.

Our paper makes the following contributions:

\begin{enumerate}
\item We created a programming model that uses a card ``game'' metaphor to orchestrate the behaviors of an autonomous physical system. The salient characteristic of this model is the use of \emph{cards} in \emph{hands} and \emph{decks} to specify concurrent and sequential behaviors. Additionally, physical hardware is abstracted through a system of \emph{token} cards, and data flows through a system of \emph{yields} and \emph{inputs}.
\item We designed a set of drone-specific cards on paper to rapidly evaluate our system. Our user testing demonstrated that our system was easy to learn, engaging, and highly expressive with acceptable levels of mental effort.
\item We created a prototype implementation of our system, called CardKit. It includes an implementation of drone-specific behaviors using the DJI API~\cite{DJI:2017}. CardKit is implemented in the Swift programming language and has been released open source at \url{http://github.com/CardKit}.
\end{enumerate}

\section{Related Work}
\label{sec:related}

\subsection{Programing Paradigms}

Several important programming paradigms, including many visual paradigms~\cite{Myers:1986, whitley1997visual} inspired our approach as we developed our card-based metaphor.

\subsubsection{Block-based Programming}

Block-based programming environments, such as as Scratch~\cite{maloney2004scratch, Malan:2007}, RAPTOR~\cite{Carlisle:2005}, and Tickle~\cite{Tickle:2017} visualize traditional programming constructs as blocks of assorted shapes, helping users learn how to structure programs. Blocks often contain visual ``connectors'' which place constraints on which blocks can be used together, making it easier to understand the relationship between blocks, such as inputs or sequenced operations.

For programming drones, block-based metaphors might be expressive enough to encompass a broad range of drone activities. However, although these tools do simplify the act of programming by helping users focus on what code means versus the syntax used to express it~\cite{Bau:2017}, fundamental sources of complexity remain: users need to learn basic programming principles -- variables, conditionals, loops, data structures, etc. -- and work in some form of lower-level ``code.'' Further, it may be challenging to program using these tools in the demanding physical environments in which drones may operate, such as in the outdoors using a smartphone or tablet device, where manipulating code-level expressions might be challenging.

\subsubsection{Dataflow Programming}

Dataflow programming~\cite{sousa2012dataflow} is a style of visual programming in which functions and operations exist as nodes in a graph, with edges representing data dependencies. Node-RED~\cite{NodeRED:2017}, Cantana~\cite{young1995cantata}, and Fabrik~\cite{Ingalls:1988} are examples of dataflow programming. It is a popular form of programming for IoT devices in which a stream of data is received, transformed, and then acted upon.

Our card-based programming model shares some similarities with the dataflow model. We allow data to flow through a card program by having data \emph{yielded} by cards accessible as \emph{input} to subsequent cards. However, the flow of data is strictly unidirectional; card programs are cycle-free DAGs, not arbitrary graphs. In addition, card programs contain a single entry point -- the first hand -- rather than an arbitrary number of data-producing sources.

\subsubsection{Trigger-Action Programming}

Trigger-action programming uses the construction of ``if some event happens, then perform some action.'' In an evaluation of this metaphor, Ur et al.~\cite{Ur:2014} found that ``this paradigm can express many of the smart-home behaviors that immediately came to our participants' minds.'' However, this model is highly \emph{reactive} and depends on triggers firing when an external event occurs in order to perform some action. For programming drones or other IoT devices that may need to \emph{actively} initiate behaviors, a pure trigger-action model may not be adequate.

We have adopted the notion of a trigger in our model by defining cards that explicitly act as triggers, waiting for an external event to occur. For example, the \textul{SetATimer} card triggers~\footnote{By \emph{triggers}, we mean that the card's end condition becomes satisfied. We discuss end conditions in the next section.} after the specified amount of time has elapsed, and the \textul{WaitUntilLocation} card triggers after a certain geolocation is reached.

\subsubsection{Card Programming}

Magic Cards~\cite{zhao2009} uses cards to implicitly define tasks a robot must perform in a physical environment. Cards are placed in different rooms to specify tasks to perform in those rooms. In a home, the robot moves around the house searching for tasks and executes those tasks when they are discovered. Cards are grouped into different categories (action, modifier, order, special purpose, and object) and some cards can modify the behavior of others. This solution works well for robots that are situated in a small physical space, as the use of the cards is highly contextual -- e.g. ``mop the floor'' is placed in the kitchen, and ``vacuum the carpet'' is placed in the bedroom. For UAVs and other robots operating in different environments -- e.g. in the air, underwater, or other places inaccessible to people -- the placement of physical cards in physical spaces to specify robot behaviors breaks down.

\subsection{Existing Drone APIs}

Popular drone manufacturers such as DJI and Parrot provide low-level APIs for controlling drones~\cite{DJI:2017, Parrot:2017}. These APIs are flexible, powerful, and well-supported with frameworks for programming languages including Swift, Python, and Java. Although these frameworks possess the broadest possible expressibility in terms of what kinds of tasks can be accomplished with a drone, writing such solutions may require substantial software engineering expertise and effort. In comparison, our card programming model reduces the possible space of expressible programs in favor of simplifying those programs that can be expressed. The degree to which complex programs are simplified is discussed later as part of the results of our user study.

\subsection{Collectible Card Games}

Our approach borrows several concepts from collectible card games (CCGs) such as \emph{Hearthstone} and \emph{Magic: The Gathering}. CCGs have a set of cards and rules that define how and when cards are played. Cards typically have standard designs depicting important information, with an icon or artwork representing the card, a description, a ``cost'' for playing the card, constraints on playing the card (e.g. prerequisites), and the outcome or effects of playing the card. Our model borrows these specific concepts from CCGs:

\begin{itemize}
\item We use the terminology of \emph{cards} to represent single behaviors or actions, \emph{hands} to represent a set of concurrent activities, and \emph{decks} to represent a sequence of operations.
\item Our use of tokens to limit concurrent behaviors is modeled after the notion of ``mana'' or ``gold'' in CCGs as a way of managing finite resources. For example, a camera token represents the fact that a drone possesses a single camera. Thus, within a hand, only one action card may be played having a dependency on the camera token.
\item Our cards contain artwork for easy identification and are color-coded based on their function.
\item Some cards are ``played'' on top of other cards, such as when they serve as inputs or modifiers.
\end{itemize}

\section{Card Programming Mechanics}
\label{sec:mechanics}

We present the details of our card-based programming model in the context of drones, as the operations of a drone are relatively simple and easily understood: a drone can \emph{fly} to a particular location, it can \emph{capture} data using a camera or other sensors, it can perform \emph{computations}, such as computing flight plans or analyzing images, and it can \emph{communicate} via feedback in a mobile app. These qualities comprise the sense, think, act, and communicate paradigm for robots~\cite{siegel2003sense}. We note that although our presentation is drone-centric, we discuss later how the model can be applied to other kinds of robots and IoT devices that are able to sense, think, act, and/or communicate.

%The structure of our programming model follows a card game with a few important exceptions. To the extent the metaphor holds, the ``game'' is a sort of solitaire or puzzle, with the goal of finding a configuration of cards to accomplish an intended task. There is no notion of scoring or competition.

Our ultimate aspiration is to create a digitized means to construct drone programs, and later in this paper we discuss our implementation effort. As a first step, we began our design process using paper cards in order to rapidly prototype and test the system, and iterate where needed.

% in the field via a mobile device (e.g., an iPad or Android app). We envision that before an actual drone flight, a drone operator assembles the drone cards and ``compiles'' them, translating them into executable machine instructions. Optimization steps can remove redundant or extraneous cards. In a full digitized form, compilation and optimization would be an iterative process with the user, helping to ensure that the cards accurately reflect user's intent prior to execution of the drone flight.

\subsection{General Structure: Cards, Hands, and Decks}

A \emph{deck} (program) consists of several sequential \emph{hands} (steps). In each step, a ``player'' (programmer) lays out a set of cards, or \emph{hand}, interpreted collectively as a unit.

\begin{figure}[ht]
\centering
\includegraphics[width=2.5in]{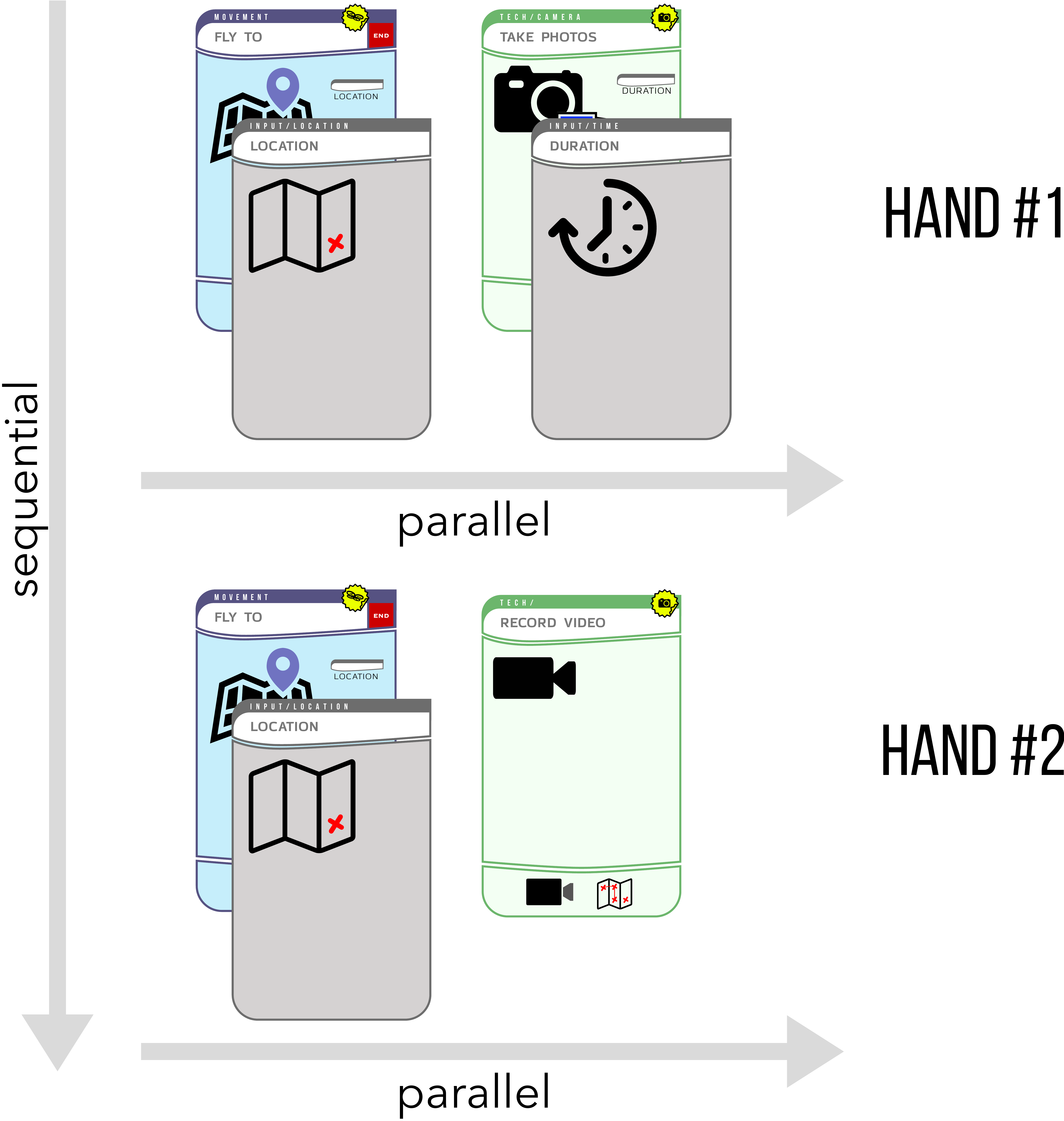}
\caption{In each hand, multiple cards represent concurrent drone activities (horizontally). Stacked cards are of type \emph{Input}. Multiple (vertical) hands represent a sequence of independent activities.
}
\label{fig:card-execution}
\vspace{-0.5em}
\end{figure}

A hand represents a set of concurrent drone behaviors; thus there is generally no defined order amongst cards in the same hand. However, there is one exception -- some cards in the hand might \emph{modify} or provide \emph{input} to other cards. In that case, the card is played by stacking it on top of the card it modifies. Figure~\ref{fig:card-execution} depicts how computation occurs sequentially across hands and in parallel within a hand.

\begin{figure}
\centering
\raisebox{-0.5\height}{\includegraphics[width=1in]{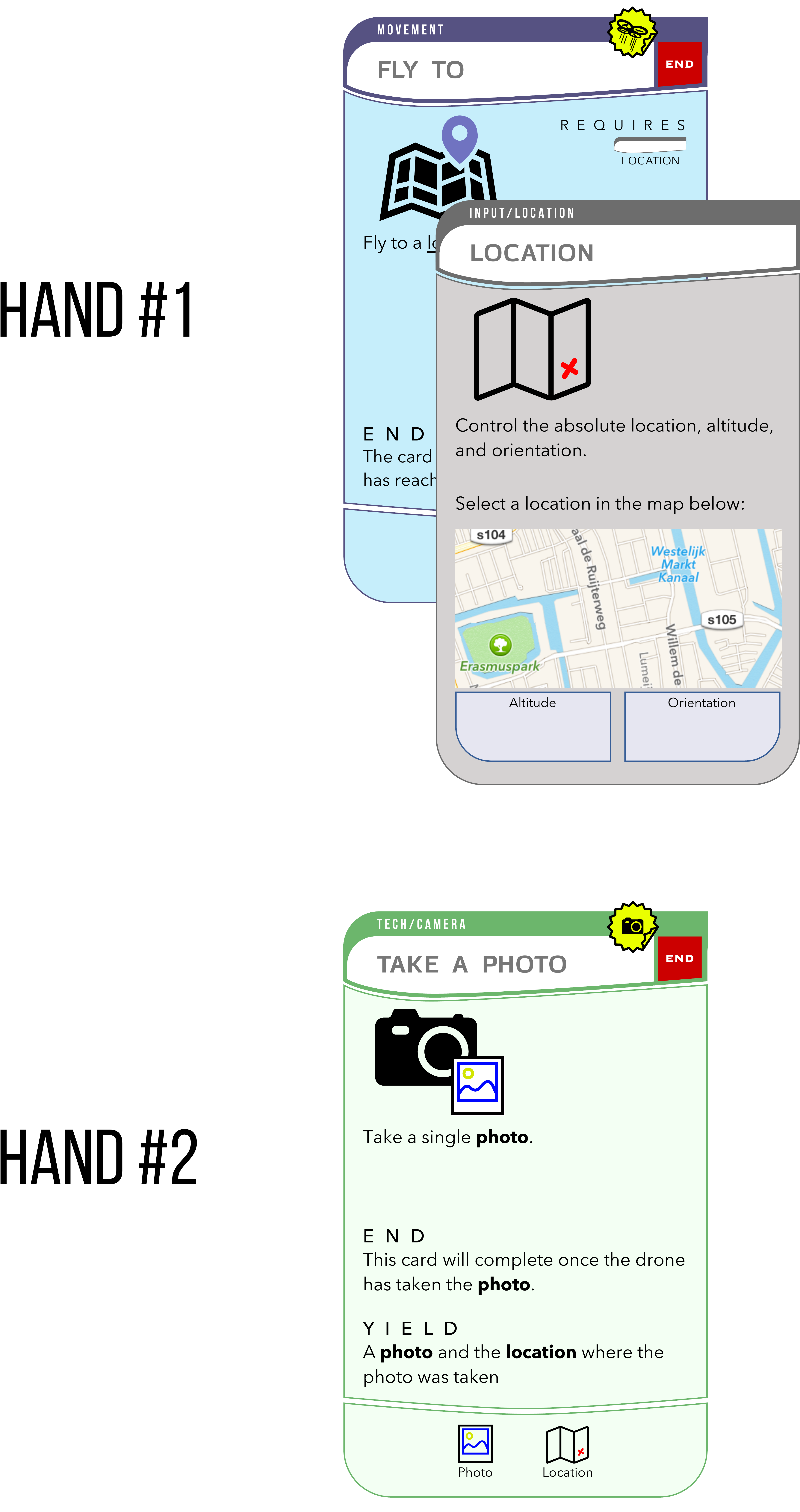}}
\hspace{0.05in}
\unskip\ \vrule\
\hspace{0.05in}
\raisebox{-0.5\height}{\includegraphics[width=2.1in]{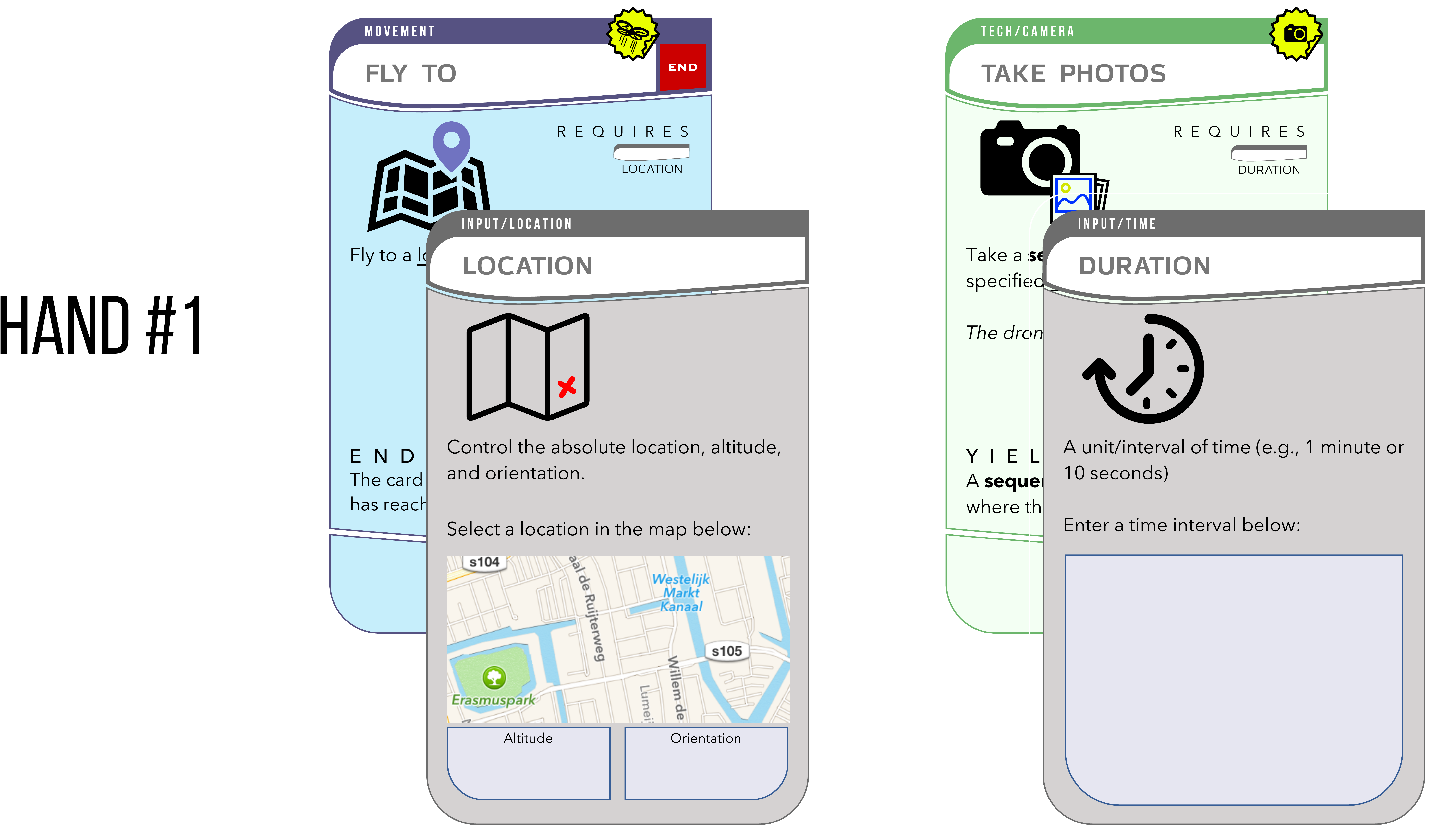}}
\vskip 1em
\caption{Sample card programs: (a) Fly to a location and take one photo. (b) Take photos while flying.}
\label{fig:example-one-two}
\end{figure}

Figure \ref{fig:example-one-two}(a) showcases a simple example drone program. It instructs a drone to fly to a location and take a photo. The first of two hands contains a \textul{FlyTo} card and a \textul{Location} card. The \textul{Location} card acts as an input to the \textul{FlyTo} card, specifying the location to which the drone should fly. The drone will first fly to that location and then proceed to the next hand. In the second hand, the \textul{TakeAPhoto} card instructs the drone to take a photo. The drone will take a photo and proceed to the next hand. In this scenario, the sequence of hands ends here and the drone program has completed. In order to provide a degree of safety, we implicitly assume the presence of a \textul{Land} card at the end of each deck to ensure the drone does not hover indefinitely after completing its mission.

All cards in a hand must complete (or ``end'') before proceeding. The \emph{end condition} is indicated with an ``End'' icon located at the top right corner of the card. If an end symbol does not exist, the card will execute indefinitely. In the example in Figure~\ref{fig:example-one-two}(a), since \textul{FlyTo} contains an end condition, execution proceeds to the next hand only when the \textul{FlyTo} card has satisfied its end condition by reaching its specified destination.

In Figure~\ref{fig:example-one-two}(b), the first half of the hand contains a \textul{FlyTo} card, which has an end condition, and the second half contains a \textul{TakePhotos} card, which does not. A \textul{Duration} card specifies a period for taking photos. Thus, the drone will fly to the specified location while taking photos. If an indefinite duration is specified, the drone will continue hovering and taking photos ``forever,'' until the battery is consumed and the drone lands itself.

\begin{figure}
\centering
\includegraphics[width=2in]{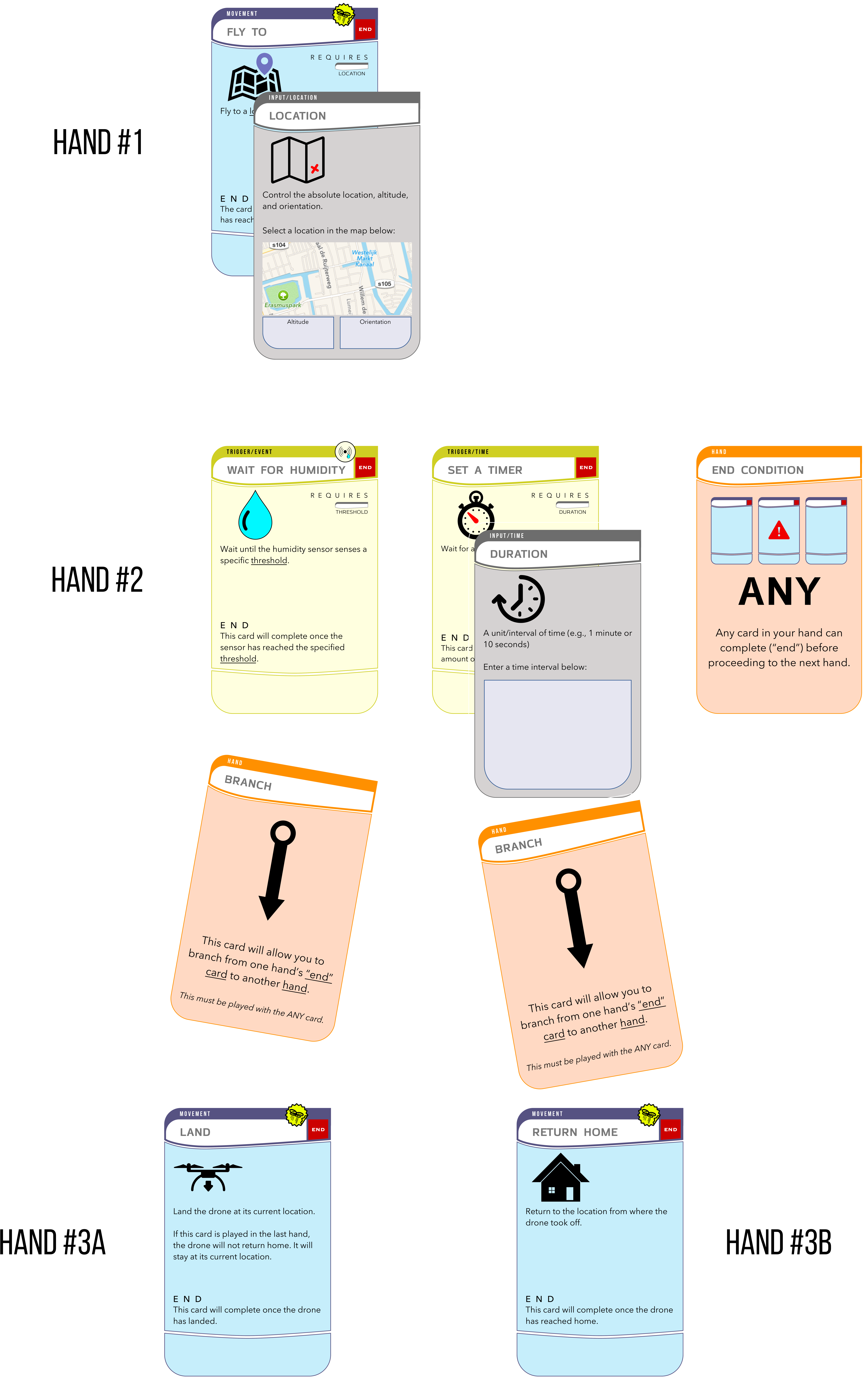}
\caption{Card program for detecting and reacting to humidity levels. Note how branching directs control flow based on which \emph{Action} card ends first.}
\label{fig:example-three}
\vspace{-0.5em}
\end{figure}

Generally, card programs contain one hand per program step. However, we support a form of conditional branching logic via \emph{branch} cards. These cards enable the drone to branch to a different hand depending on which \emph{Action} ends first. Figure~\ref{fig:example-three} shows an artificial example in which a drone either lands when a desired humidity level is measured, or returns home after a certain amount of time has elapsed.

%In Figure \ref{fig:study-example3}, we show an example in which the drone will perform different behaviors depending on which \emph{Action} ends first -- either an object is detected and then followed, or a timer fires and the drone returns home.

%Generally, card programs contain one hand per program step. However, we support a form of conditional branching logic via \emph{branch} cards. In Figure \ref{fig:example-three}, the second hand contains two sub-hands representing the branching logic. In this program, the drone will fly to a location and either wait for a specific humidity reading or wait for 10 minutes. Whichever event occurs first triggers the step to the next hand. If the humidity reaches a defined threshold, the drone will land at that location (Hand 3A). If the 10 minute timer expires before the drone detects the humidity threshold, the drone will return home (Hand 3B).

\subsection{Card Details}

\begin{figure}[tb]
\centering
\includegraphics[width=2.25in]{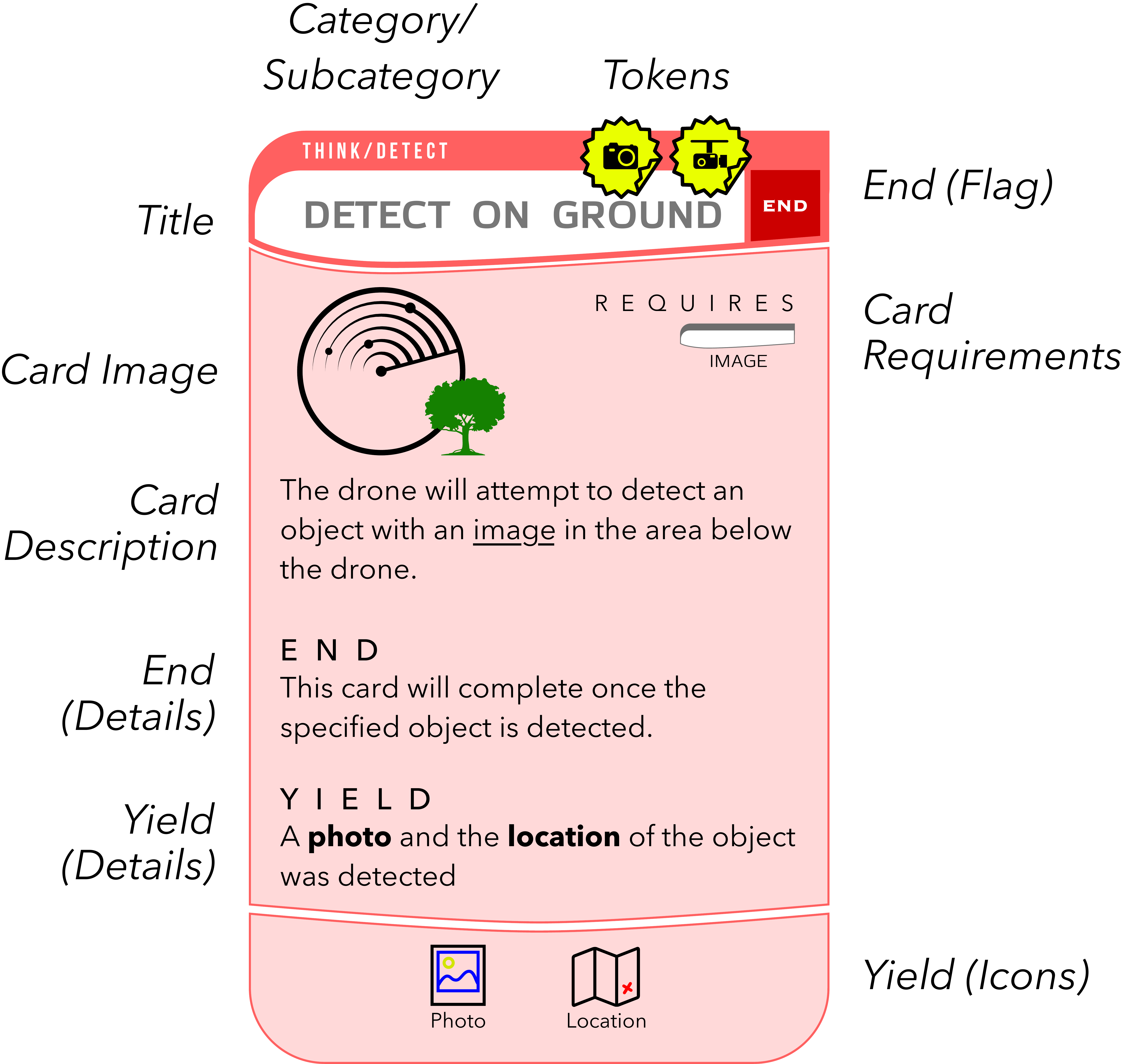}
\caption{Example card \textul{DetectOnGround} showing all possible elements: tokens, end condition, requires, end details, yield details, and yield icons. These elements are optional and will not appear on all cards.}
\label{fig:card-layout}
\end{figure}

Figure~\ref{fig:card-layout} illustrates an example card design. The colored background (red) indicates the card class (\emph{Think}). At the top-left, the card class and subclass are given with dark background (\emph{Think/Detect}); below is the name of the card in large print (``Detect On Ground''). At the top-right, the yellow starbursts indicate the tokens consumed by this card (camera and gimbal), and the red ``End'' indicates that the card imposes an end condition. The curved design of the top header is intended to evoke the shape of a drone's propeller. In the middle section, every card has a unique card image for quick reference. To the right of image, we depict any prerequisite cards required for this card to be played, presented by color-coded class (a single \emph{Image} card). Below, the card describes its purpose, the requirements of its end condition (if applicable), and the data it produces (called \emph{yields}).

We categorize cards by function into one of five classes.

\begin{enumerate}
\item \emph{Action}: Causes the drone hardware to perform some behavior. Example: \textul{FlyTo} commands the drone to go to the GPS coordinate specified by an \emph{Input} card.
\item \emph{Input}: Specifies a numerical, categorical, textual, or other data input used by another card. Example: \textul{Location} represents a GPS coordinate and is used by \textul{FlyTo}.
\item \emph{Hand}: Modifies execution behavior of the entire hand. Example: \textul{Any} makes a hand require only one end condition to be satisfied before execution proceeds to the next hand.
\item \emph{Deck}: Modifies the execution behavior of the deck. Example: \textul{RepeatDeck} repeats execution of the deck from the first hand after all hands have completed. Presently, this is the only \emph{Deck} card we have defined, but future \emph{Deck} cards can exert global rule modifications across all hands (e.g. speeding up or slowing down all movements, setting resolution options for the camera, etc.).
\item \emph{Token}: Grants access to underlying hardware. Starburst tokens are consumed (usable by only one card in a hand), circular tokens are not. Example: \textul{Movement} is used to control a drone's movement, and \textul{Camera} provides access to the camera.
\end{enumerate}

In our design, \emph{Action} cards have also been split up across a number of subclasses according to their function.

\begin{enumerate}
\item \emph{Movement} cards control the movement of the drone. Example: \textul{FlyTo} causes the drone to fly to the specified location.
\item \emph{Tech} cards operate a hardware token. Example: \textul{TakePicture} captures a single photo.
\item \emph{Think} cards perform computational tasks. Example: \textul{DetectInAir} applies computer vision on an \textul{Image} to recognize nearby flying objects. Note that we do not specify how this computational task actually occurs, whether on the drone, on a ground station, or via a cloud service; we expect the specific implementations of these cards to determine their own implementation requirements.
\item \emph{Trigger} cards are used for asynchronously-satisfied end conditions. Example: \textul{SetATimer} becomes satisfied only after a specified amount of time has elapsed.
\end{enumerate}

To aid discovery and recognition, cards of the same class, or subclass in the case of \emph{Action} cards, are printed in the same color.

Cards from the \emph{Input} class affect the execution of a single card, on top of which it is stacked. At a higher level, \emph{Hand} cards may affect execution of multiple cards in one hand or the transition from one hand to the next. \emph{Deck} cards affect execution across multiple hands.  Otherwise, all cards within a hand are fully independent.

\subsubsection{End Conditions and Branching}

A hand is not complete until its \emph{end conditions} are satisfied. Some cards (but not all) are marked with ``End.'' These cards complete a finite amount of work prior to completion. For example, \textul{TakeAPhoto} is marked as the card will complete as soon as the photo has been captured. Contrastingly, \textul{RecordVideo} does not complete -- it will keep recording video for an arbitrary amount of time, so long as the hand continues. By default, a hand ends when all end conditions in the hand are reached. This logic may be reversed using the \textul{Any} card; when this card is played in a hand, once any card satisfies its end condition, it triggers the end of the hand and the movement to the next hand. Generally, all hands should have at least one card with an end condition. Hands in the last step -- the ``final'' hand -- are not required to have an end condition, as they will continue indefinitely until the drone's battery reaches a critical threshold, triggering it to immediately land.

The \textul{Branch} enables logical branching, depending on which end condition is satisfied first. Using the \textul{And}, \textul{Or}, and \textul{Not} cards, arbitrary Boolean logics can be created to determine when a hand ends and which hand should be executed next. For example, given \emph{Action} cards A, B, C, and D, the following hand can be constructed: $(AND(A, B)\ ;\ \textrm{Branch}(2a))\ ;\ (OR(C, D)\ ;\ \textrm{Branch}(2b))$. This hand will branch to hand $2a$ when both $A$ and $B$ have ended, or it will branch to hand $2b$ when either $C$ or $D$ have ended, whichever occurs first.

\subsubsection{Side Effects and Yields}
Drone activities have real-world consequences. For example, an item might be picked up and flown from one location to another. We do not model these kinds of physical side effects, but we do model side effects that result in data captured by the drone. Cards may have one or more \emph{yields}. For example, the \textul{TakePhotos} card yields a sequence of photos and a sequence of locations at which the photos were taken. In our design, yields are prominently presented at the bottom on each card so that one can quickly check for desired data capture.

During design and early testing, we debated whether yields should be reusable across game steps. Might the yield produced in one hand be consumed as an input to a later hand? While technologically feasible due to a strict adherence to strong typing, and perhaps desirable due to its added expressiveness, we felt this feature would add significant complexity and mental effort. Thus, we initially treated yields only as a persistent side effect -- data yielded by the drone would be available for analysis only upon completion of program execution. However, during our evaluation, participants asked to reuse yields across hands. Therefore, our implementation of yields in CardKit does enable yields produced in prior hands to be consumed as inputs in downstream hands.

\begin{table*}
{
\scriptsize
\sffamily
\begin{tabularx}{\linewidth}{cXp{3cm}p{3cm}}
\textbf{Symbol} & \textbf{Meaning} & \textbf{Example}  & \textbf{Explanation} \\ \hline \hline
$\leftarrow$    & Denotes \emph{Input} relationships & \textul{CoverArea} $\leftarrow$ \textul{BoundingBox} & \textul{BoundingBox} is an input to \textul{CoverArea} \\
$()$ and $+$    & Specifies multiple inputs to a card & \textul{Follow} $\leftarrow$ (\textul{RelativeToObject} + \textul{TrackOnGround} $\leftarrow$ \textul{Image}) & \textul{Follow} requires a Relative card and a Track card as inputs. \textul{TrackOnGround} also requires an \textul{Image} as input \\
$(\#)$ & Indicates parameters to a card or identifies a branch & \textul{Repeat}(3) & Repeats the hand three times \\
$;$             & Separates cards in a hand & \textul{FlyTo} $\leftarrow$ \textul{Location} ; \textul{TakePhotos} & Drone will fly to a specified location and take photos at the same time \\
$\{\}$          & Groups cards belonging to the same branch & (\textul{FlyTo} $\leftarrow$ \textul{Location} ; \textul{Branch}(1)) ; (\textul{DetectOnGround} $\leftarrow$ \textul{Image} ; \textul{Branch(2))} & Branch 1 will be followed if the drone arrives at the location. Branch 2 will be followed if the drone detects the input image. \\
$[]$            & Specifies the kind of information provided to an \textul{Input} card & \textul{Location} [\textit{my house}] & The location used for input is ``my house''. \\
\end{tabularx}
}
\vspace{-1em}
\caption{Syntax for card solutions\label{tbl:syntax}}
\vspace{-1.5em}
\end{table*}

\subsubsection{Composition of a Single Hand and the Token System}

In CCGs, there is often some notion of cost. Cost imposes limits and tradeoffs on what the player might be able to accomplish in the same hand. \emph{Hearthstone} lets players cast spells, but only in number and strength as their casting costs allow with the current level of available mana. \emph{Dominion} only allows players to buy a certain number of cards per turn based on available gold. We build on these models to express and enforce limits on what the drone might do concurrently, without requiring the user to explicitly understand any computer science concurrency idioms.

A single hand is composed of one or more cards. Although there are no specific limits to the number of cards in any one hand, there are some constraints on the number of times any particular card might be played and with which other cards it might be played. We represent these constraints through \emph{tokens}, depicted as a starburst with an icon. In assembling a deck, the user adds a set of \emph{Token} cards corresponding to physical characteristics of the drone on which the program will be run. Each \emph{Token} card represents a single hardware capability of the drone. For example, a flying drone would have a single \textul{Movement} token. If the drone possessed a camera mounted to a multi-axis gimbal, the player would also add \textul{Camera} and \textul{Gimbal} tokens to their deck. Within a single hand, a token may be \emph{consumed} at most once. For example, all \emph{Movement} cards consume a \textul{Movement} token. Hence, it would be disallowed to play \textul{FlyTo} and \textul{Circle} in the same hand as they both require the \textul{Movement} token. They could, however, be played in two successive hands.

Some tokens are not consumed upon use. These tokens define a hardware aspect of a drone that \emph{can} be used concurrently. In our design, these tokens are represented as a circle rather than starburst (a circle is like a starburst with infinite points, reflecting that such cards may be used an arbitrary number of times). For example, the \textul{Humidity Sensor} produces a stream of humidity values that can be consumed by multiple cards, such as \textul{LogHumidity}, which logs the humidity value, and \textul{WaitForHumidity}, which ends when the humidity reaches a given threshold.

\subsubsection{Computational Play}

In its paper form, users lay out a sequence of hands row by row on a flat surface. This layout specifies a program, which is read in sequence from top to bottom. As each hand is treated as an unordered collection of cards, left-to-right orderings may be disregarded.
Figure~\ref{fig:card-execution} illustrates a representative layout of cards in hands.

In advanced scenarios, the program may incorporate conditional branching logic. In such cases, two or more hands may be considered as alternatives in the same step. We suggest each alternative hand be laid out horizontally with gaps to clearly indicate which subsequent hands branch from which end conditions in the previous hand.

\section{Evaluation}

Our evaluation consists of a paper prototype user study. We consider three primary metrics: ease, expressiveness, and engagement. Ease represents how much difficultly users perceive in selecting and laying out cards to accomplish their programming goal. We presented participants with specific tasks and evaluated their performance using several objective metrics such as accuracy, completion time, and established scales of cognitive load and usability. Expressiveness represents the diversity of tasks that can be programed. We asked participants to propose and implement a novel, creative drone task. Engagement represents the extent to which our format yields an enjoyable programming experience. Using Likert-format and free response metrics, we evaluate whether participants enjoyed our whimsical, game-like structure. We recruited 18 participants for an in-person, one-on-one, two-hour evaluation session at our facility. Table~\ref{tbl:demographics} shows a description of our participants. Our study protocol was reviewed and approved by our institution's human subjects research review board.

\begin{table}[t]
{
\sffamily
\scriptsize
\centering
\vspace{.5em}
\begin{tabularx}{\linewidth}{lX}
\textbf{Question}   & \textbf{Participants (\# / \%)}     \\ \hline \hline
Gender & Male (\textbf{11 / 61\%}), Female (\textbf{7 / 39\%})  \\
Age Range & $\leq$20 (\textbf{2 / 11\%}), 21-35 (\textbf{15 / 83\%}), 36-50 (\textbf{1 / 6\%})\\
\% time writing code &  $\leq$20 (\textbf{5 / 28\%}), 21-40 (\textbf{3 / 17\%}), 41-60 (\textbf{5 / 28\%}), 61-80 (\textbf{3 / 17\%}), $\geq$81(\textbf{2 / 11\%})\\
Experience with \ldots & \\
\ldots Flying drones & Yes (\textbf{3 / 17\%}), No (\textbf{15 / 83\%}) \\
\ldots Visual programming & Yes (\textbf{4 / 22\%}), No (\textbf{14 / 78\%}) \\
\ldots Card games & Yes (\textbf{6 / 33\%}), No (\textbf{12 / 67\%})
\end{tabularx}
}
\vspace{-.5em}
\caption{Demographics and pre-study questionnaire.}
\label{tbl:demographics}
\vskip -1em
\end{table}

A member of our research team conducted the evaluation sessions. During these sessions, the evaluator began by explaining the format and goals of the study and received informed consent from the participant. The evaluator then explained the basic rules via a short booklet and laid out all of the paper cards in stacks on a table in front of the participant. Explanations of rules and mechanics were kept consistent across participants to the extent possible.

In the training phase, the evaluator demonstrated three example tasks and their solutions. Next, in the evaluation phase, participants performed three experimental tasks on their own, with the evaluator providing feedback only in the case of syntactic rule violations. Each of these tasks were presented in a randomized order so that we may evaluate task difficulty independent of learning effects. When a participant made rule violations or other ``syntactical'' errors during the evaluation phase (i.e., errors that a real compiler could automatically detect and report in a digital format), we disclosed and explained these errors to the participant. No other feedback or hints were given during the evaluation phase; especially, logic errors were not disclosed. The accuracy of each participant's solution was scored against a pre-defined rubric.

After each task, participants completed the the NASA Task Load Index (TLX)~\cite{hart1988development} and were asked questions regarding their experience (e.g., how enjoyable or hard). The TLX measures perceived workload across six sub-scales: mental, physical, temporal (time pressure), performance, effort, and frustration. In our study, we applied it to understand the difficulty of each individual evaluation scenario. We also recorded the amount of time spent on each task and took a picture of the final outcome. At the end of the study, participants filled out a final questionnaire and completed the System Usability Scale (SUS)~\cite{brooke1996sus} to quantify effectiveness, efficiency, and satisfaction for the overall approach. Time and number of cards required per task serve as additional metrics of task complexity. For their time, participants received a \$10 lunch voucher and the opportunity to fly a small consumer-grade drone indoors.

\subsection{Drone Scenarios}

We constructed six drone scenarios for our study, three for instruction and three for testing. For the training scenarios, participants were shown the solution and given an explanation as to how the solution operated. For brevity, we show only the three testing scenarios below. Table~\ref{tbl:syntax} gives an overview of the syntax we use to present solutions in this paper; participants were shown solutions using the actual paper cards.

For the evaluation tasks, we provide our rubric to quantify the accuracy of participant answers. Since there is no single correct solution for any scenario, solutions will vary in number of cards, number of hands, and complexity. Participants' solutions were scored only for what they did correctly; no points were deducted for extraneous functionality.

\noindent \textit{\textbf{Task A -- Package Delivery}: A customer has put in a request for a package to be picked up from her house and dropped off at a friend's house. The drone will go to the customer's house and land. The drone will wait there until the customer has placed the package into the drone's claw. The customer will push the button when everything is ready. Once the button is pushed, the claw will close and the drone will fly to the friend's house. Next, the drone will drop the package off from a safe height (0-5 feet above the ground). After the package is dropped off, a sound will be played to indicate the package has been delivered. Return back home once finished.}

\begin{small}
Hand 1: \textul{FlyTo} $\leftarrow$ \textul{Location} [\textit{pickup}]

Hand 2: \textul{Land} ; \textul{WaitForButtonPush}

Hand 3: \textul{CloseClaw}

Hand 4: \textul{FlyTo} $\leftarrow$ \textul{Location} [\textit{delivery}]

Hand 5: \textul{HoverToAltitude} $\leftarrow$ \textul{Distance} [\textit{5 ft.}]

Hand 6: \textul{OpenClaw}

Hand 7: \textul{PlayAudio} $\leftarrow$ \textul{Audio} [\textit{success}]

Hand 8: \textul{ReturnHome}
\end{small}

\noindent {\sffamily\small Scoring rubric: go to customer's house (5), land (3), wait for button push (5), pick up package / close claw (5), go to friend's house (5), drop off package (5) from a safe height (3), play a sound (3), return home (5). 39 possible points.} \\

\noindent \textit{\textbf{Task B -- Skiing Coverage}: You are a journalist covering the skiing portion of the Olympic Games. You would like to record every skier that goes down the slope. The drone is at the top of the slope. The drone uses the camera to locate the first skier (nearby). The drone will record video while the skier goes down the slope. Once the skier has reached the bottom of the slope, the drone will stop following (and recording). At all times, the drone should remain above the tree line (300 feet). Upon reaching the bottom, the drone should return to the top of the slope and repeat for all subsequent skiers.}

\begin{small}
Hand 1: \textul{Hover} ; \textul{DetectOnGround} $\leftarrow$ \textul{Image} [\textit{skier}]

Hand 2: \textul{Follow} $\leftarrow$ (\textul{RelativeToObject} + \textul{TrackOnGround} $\leftarrow$ \textul{Image} [\textit{skier}]) ; \textul{Altitude} [\textit{300 ft.}] ; \textul{WaitUntilLocation} $\leftarrow$ \textul{Location} [\textit{bottom of slope}] ; \textul{RecordVideo}

Hand 3: \textul{ReturnHome} ; \textul{Altitude} [\textit{300 ft.}]

Hand 4: \textul{RepeatDeck}
\end{small}

\noindent{\sffamily\small Scoring rubric: detect skier (5), follow skier (5), record video (5), stay above tree line (3), stop at bottom of slope (5), repeat (5). 28 possible points.} \\

\noindent \textit{\textbf{Task C -- Gas Detection}: You are a gas technician and a customer has called you for a gas problem. You need to find the gas leak in the area around the house to start working on the issue. You would like to use a drone to locate the gas leak. The drone will fly around the house and search for a higher than normal gas level. If this is found, the drone will wait there for 3 minutes while playing an alarm. After the 3 minutes are over, the drone will land. If a gas leak has not been detected, the drone will return back home.}

\begin{small}
Hand 1: \{\textul{CoverArea} $\leftarrow$ (\textul{BoundingBox} [\emph{yard}] + \textul{Avoid} $\leftarrow$ \textul{BoundingBox} [\emph{house}]) ; \textul{Branch}(A)\} ; \{\textul{WaitForGas} $\leftarrow$ \textul{Threshold} [\textit{high}] ; \textul{Branch}(B)\} ; \textul{Any}

Hand 2 Branch A: \textul{ReturnHome}

Hand 2 Branch B: \textul{SetATimer} $\leftarrow$ \textul{Duration} [\textit{3 min.}] ; \textul{PlayAudioLoop} $\leftarrow$ \textul{Audio} [\textit{alarm}]

Hand 3: \textul{ReturnHome}
\end{small}

\noindent {\sffamily\small Scoring rubric: survey area (5), detect gas (5), wait (3), play alarm (3), land (5), branching logic (10), return home if no leak (5). 36 possible points.} \\

\noindent \textit{\textbf{Task D --- Create Your Own Scenario:} Define and implement any task of your choosing. If required, you may define new cards (the moderator will help ensure that another card does not already capture identical functionality).}

\subsection{Ease of Use}

\begin{figure}[htb]
\centering
\includegraphics[width=.85 \linewidth]{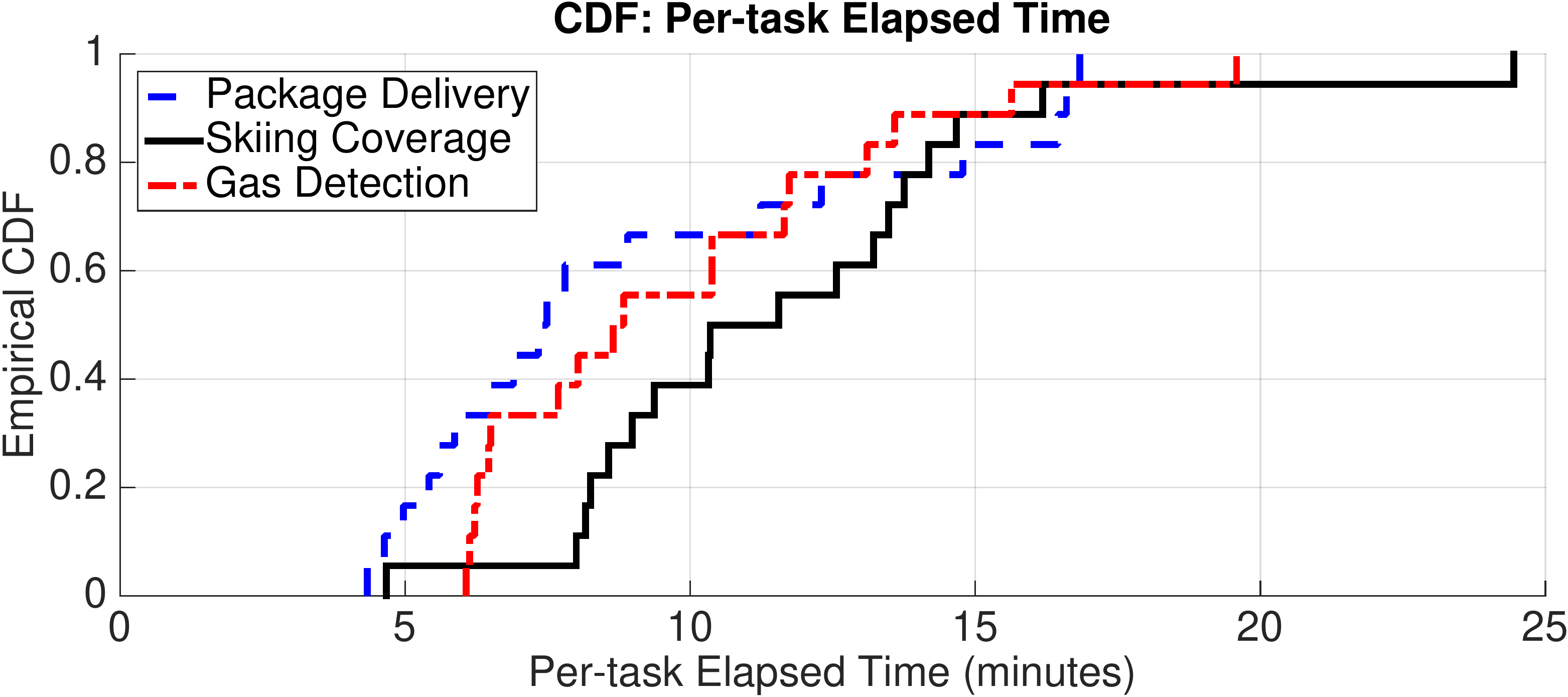}
\caption{Cumulative distribution (CDF): time elapsed, by task. There was no significant difference in completion times across scenarios, F(2,48) = 0.44, p=n.s.}
\label{fig:by_task_time_taken_cdf}
\end{figure}

\begin{figure}[htb]
\centering
\includegraphics[width=.85 \linewidth]{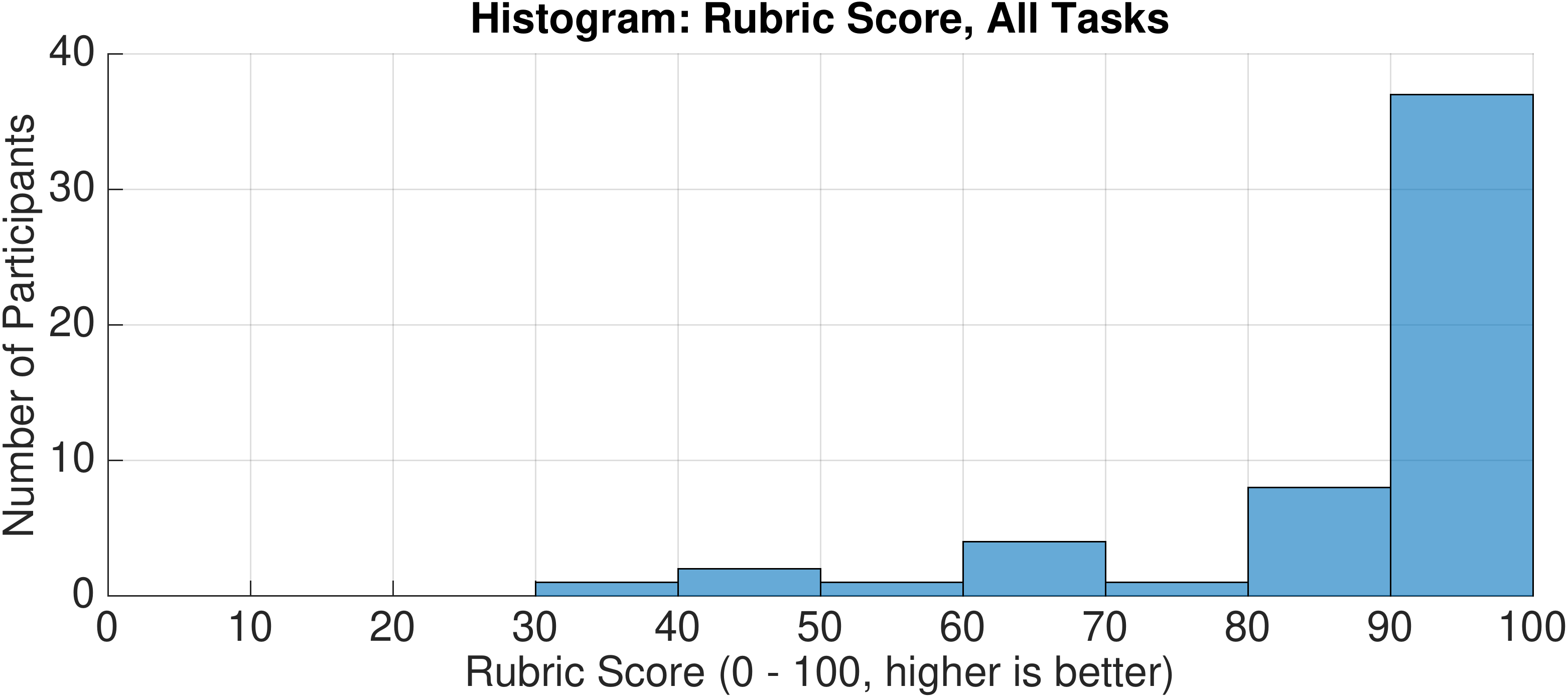}
\caption{Histogram: Normalized rubric score. M (SD) = 89 (16) points.}
\label{fig:rubric_score_histogram}
\end{figure}

Irrespective of correctness, participants used a similar number of cards as in our own solutions. Participants used the fewest cards for \emph{Skiing} (median = 12), despite tending to spend the most time on that scenario (see Figure~\ref{fig:by_task_time_taken_cdf}, a CDF of time elapsed by task). Figure~\ref{fig:rubric_score_histogram} presents a histogram of rubric scores across all tasks. Overall, participants implemented an advanced drone scenario with about 15 cards in about 12 minutes, performing the task with a high level of accuracy (M=89\% correctness) with minimal training.

\begin{figure}[htb]
\centering
\includegraphics[width=.85 \linewidth]{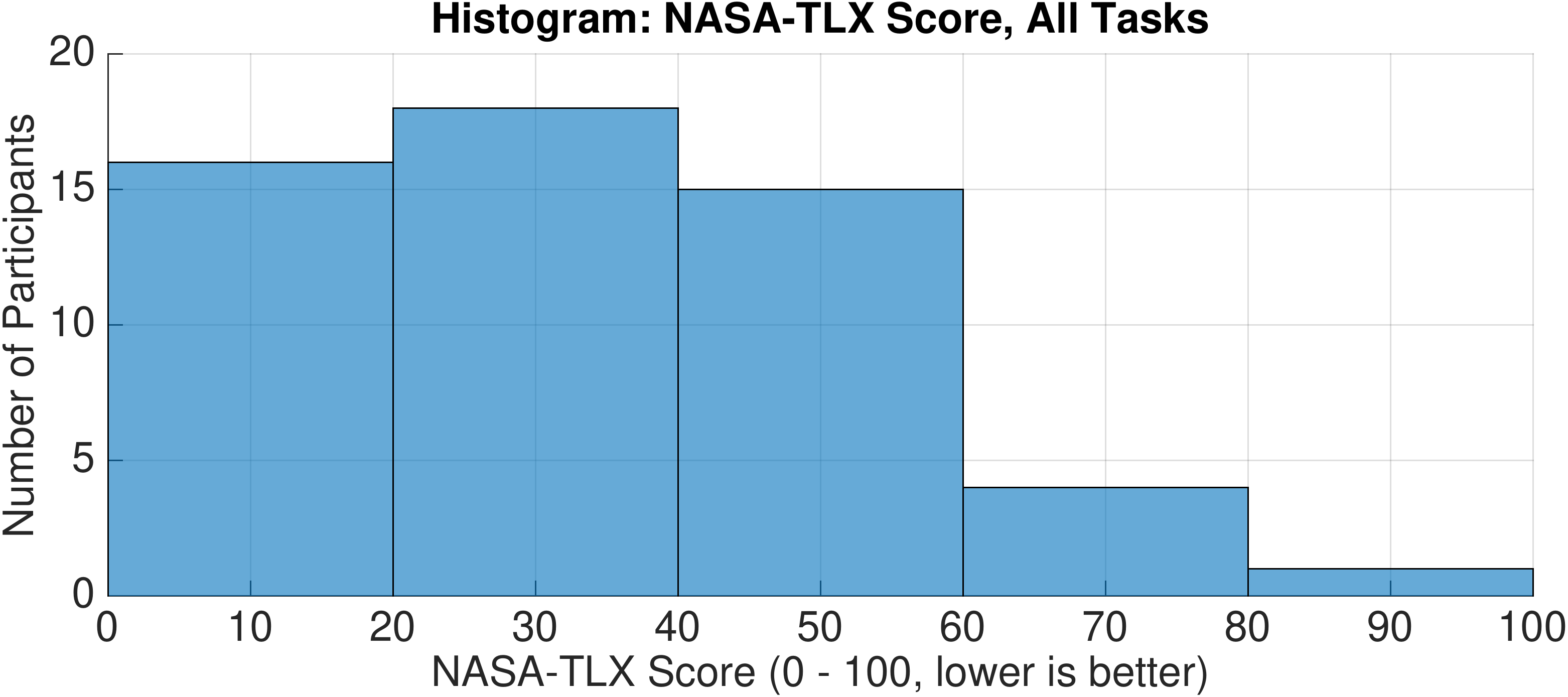}
\vskip 0.4em
\includegraphics[width=.75 \linewidth]{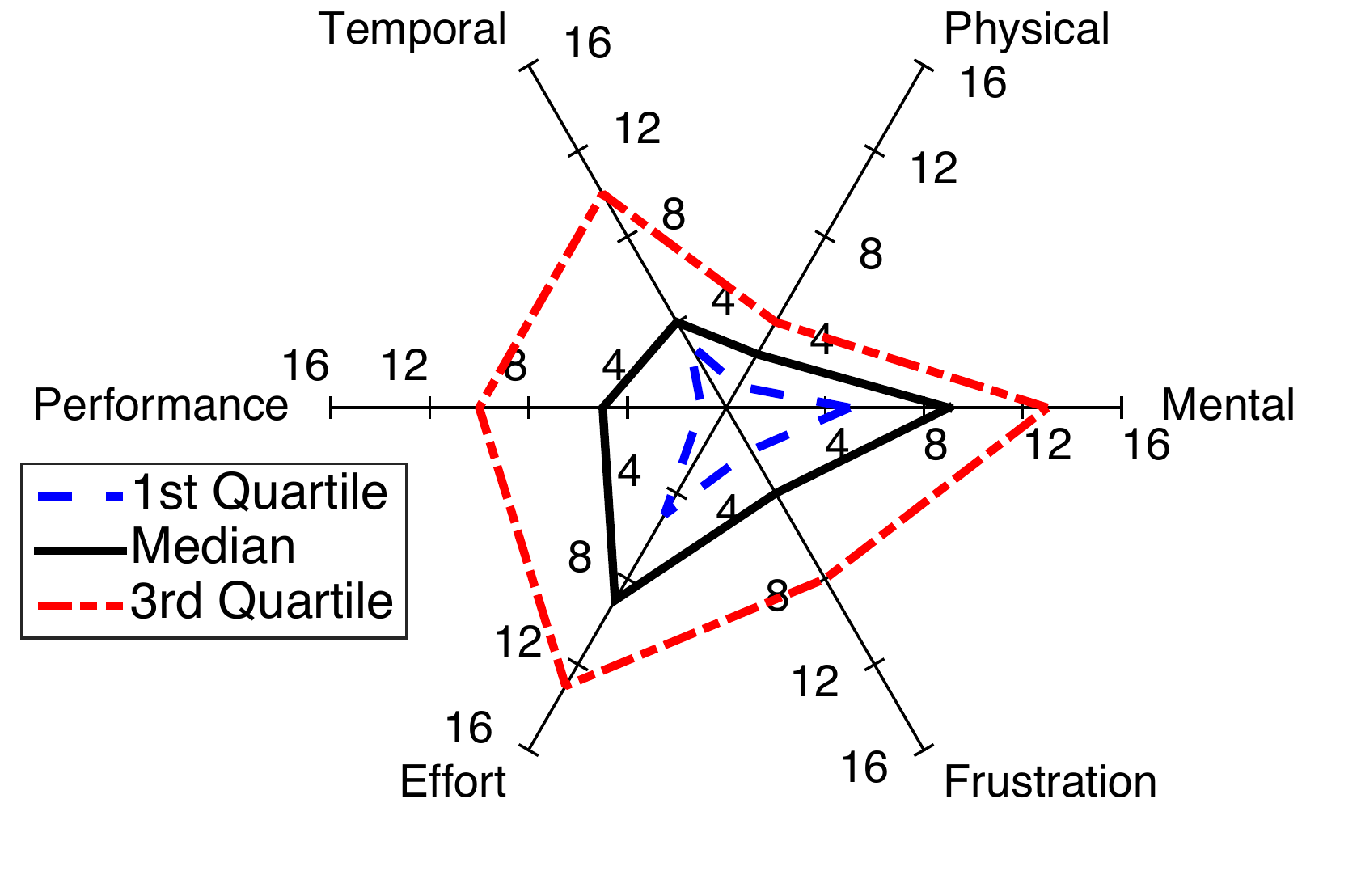}
\caption{(a) Histogram: TLX score. M (SD) = 33 (18) points. (b) Radar: TLX score (all tasks), by subscale.}
\label{fig:tlx_all}
\vspace{-1em}
\end{figure}

Figure~\ref{fig:tlx_all}(a) presents a histogram of TLX scores across all tasks (lower is better). Figure~\ref{fig:tlx_all}(b) decomposes the TLX scores along the six sub-scales. As we did not inform participants of their mistakes between tasks, there were no significant differences in rubric score across task scenarios, F(2,48) = 0.82, p=n.s.

\begin{figure}[tb]
\centering
\includegraphics[width=.85 \linewidth]{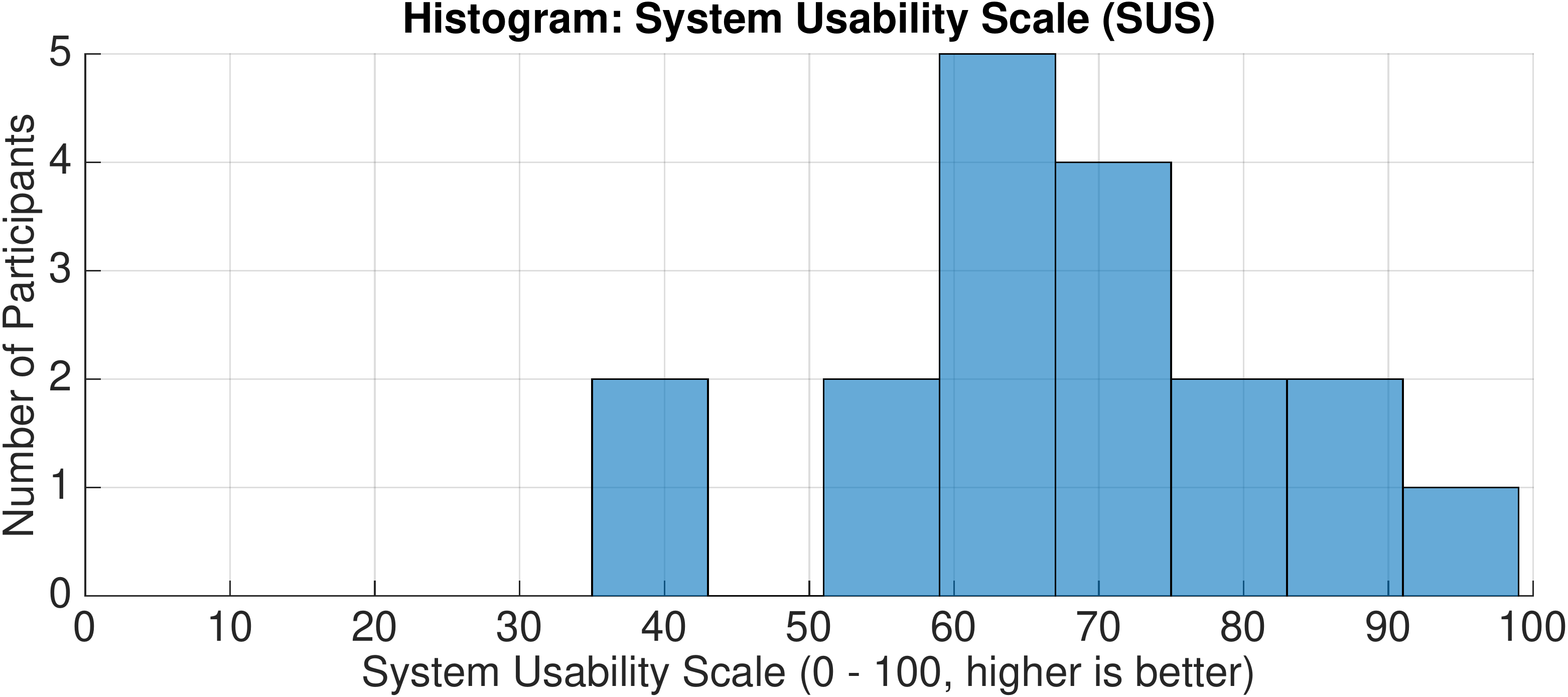}
\vskip 0.4em
\includegraphics[width=.85 \linewidth]{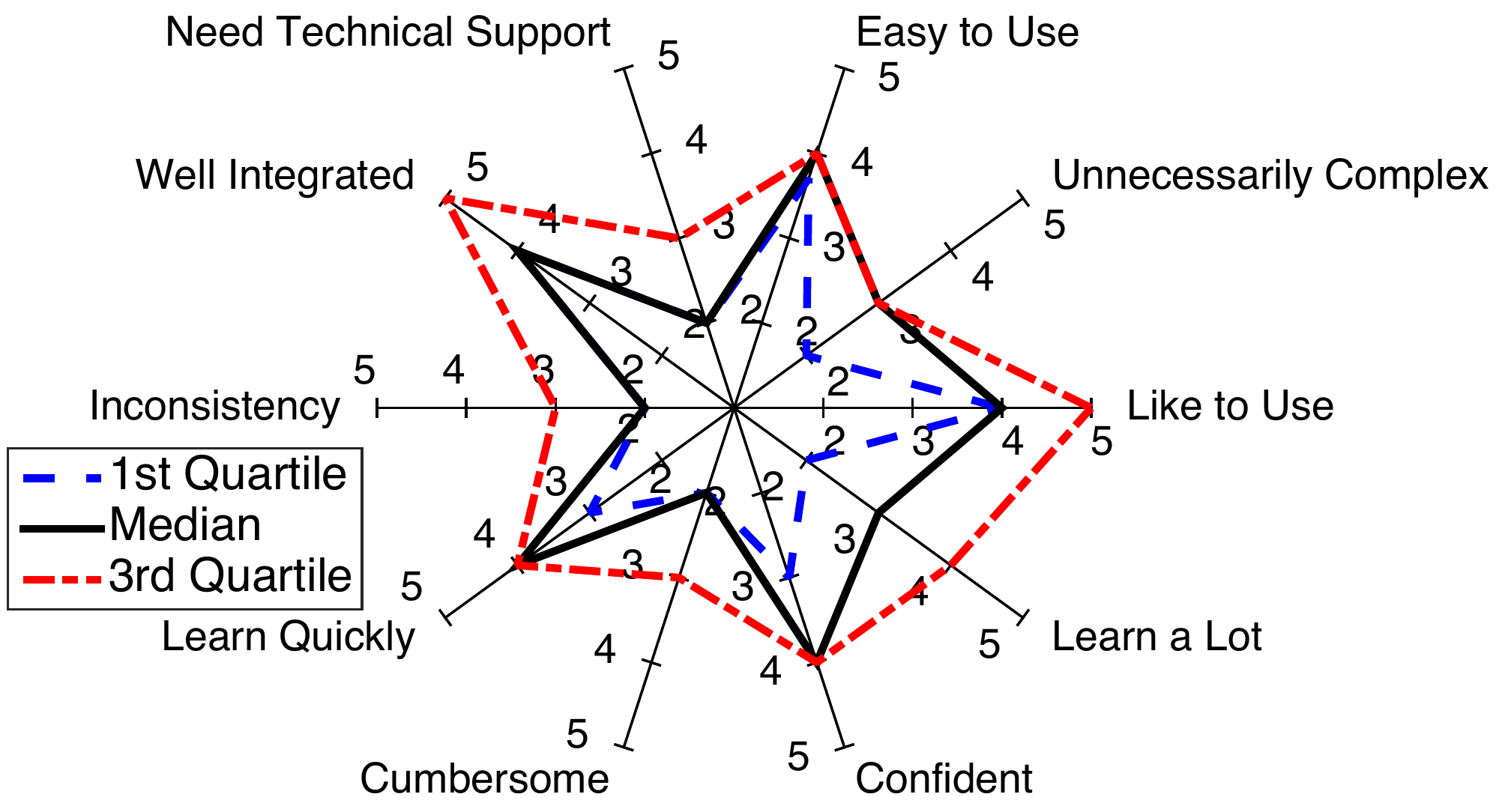}
\caption{(a) Histogram: SUS score. M (SD) = 67 (15) points. (b) SUS score, by subscale. 1 = Strongly Disagree, 2 = D, 3 = N, 4 = A, 5 = SA.}
\label{fig:sus_histogram}
\end{figure}

Figure~\ref{fig:sus_histogram}(a) presents a histogram of SUS scores. Figure~\ref{fig:sus_histogram}(b) decomposes the SUS scores along the ten sub-scales. We find that participants rated our system's usability positively (M (SD) = 67\% (15\%)), but qualitative feedback shows room for improvement (Table~\ref{tbl:post_free}). In addition, we expect some depression of SUS scores due to novelty effects and the moderate levels of Mental Demand and Effort shown in the TLX (Figure~\ref{fig:tlx_all}(b)).

\begin{table}[htb]
{
\sffamily
\scriptsize
\centering
\begin{tabularx}{\linewidth}{Xp{6.5cm}}
\textbf{Question}   & \textbf{Participants (\%)} \\
\hline \hline
Overall, how do you feel about the drone cards?	&	V. positive (\textbf{44\%}), Generally positive (\textbf{50\%}), Mixed (\textbf{0}), Generally negative (\textbf{6\%}), V. negative (\textbf{0}) \vspace{0.1cm} \\
Overall, how effective do you think the drone cards are for implementing drone scenarios? \vspace{0.2cm}  & Excellent (\textbf{17\%}), Very good (\textbf{61\%}), Good (\textbf{17\%}), Fair (\textbf{6\%}), Poor (\textbf{0}) \vspace{0.1cm} \\
Overall, how enjoyable was it to implement drone scenarios with the cards? & Very enjoyable (\textbf{50\%}), Generally enjoyable (\textbf{39\%}), Somewhat enj. (\textbf{11\%}), Very unenjoyable (\textbf{0}) \vspace{0.1cm} \\
Overall, how satisfied were you with the cards? & Very satisfied (\textbf{44\%}), Generally satisfied (\textbf{50\%}), Somewhat dissatisfied (\textbf{6\%}), V. dissatisfied (\textbf{0})\vspace{0.1cm} \\
To what extent do you feel that this system can be used to express a wide variety of drone scenarios? & A great deal (\textbf{50\%}), Some (\textbf{50\%}), Only a few (\textbf{0}), Not many (\textbf{0}) \\
\end{tabularx}
}
\caption{Post-study questionnaire. Responses were significantly correlated (Cronbach's $\alpha$ = .78), forming a single scale of efficacy, M (SD) = 61\% (11\%).}
\label{tbl:post}
%\vspace{-0.5em}
\end{table}

In the post-study questionnaire, we asked participants to rate their feelings toward the drone cards (positive or negative), the effectiveness of the cards in implementing the scenarios, their enjoyment in using the cards, their overall satisfaction with the cards, and the extent to which they felt the cards could be used to implement a wide variety of scenarios (Table~\ref{tbl:post}). These items were all rated on Likert scales. The responses to these items were significantly correlated and thus averaged to form a single scale of efficacy (Cronbach's $\alpha$ = .78). Overall, normalized efficacy was rated moderately (M (SD) = 61\% (11\%)).

Table~\ref{tbl:post_free} summarizes manually-coded free responses from the post-study questionnaire. Participants generally found the cards effective for implementing drone scenarios. In free response, a majority of participants found the cards intuitive (72\%), and many complemented the visual aspects of our design (39\%). Several criticized the quantity of cards (33\%) and complexity of the rules (22\%). Control flow and movement were most commonly cited for confusion.

\begin{table}[tb]
{
\sffamily
\scriptsize
\centering
\begin{tabularx}{\linewidth}{Xrr}
\textbf{What did you like about this card methodology?} & \textbf{\#} & \textbf{\%} \\
\hline \hline
Intuitive (e.g., easy to learn, simplifies complex tasks) & \textbf{13} & \textbf{72\%} \\
Visual aspects of design (e.g., iconography, color coding) & \textbf{7} & \textbf{39\%} \\
Systematic / consistent & \textbf{3} & \textbf{17\%} \\
Expressiveness & \textbf{2} & \textbf{17\%} \\
Fun / playful approach & \textbf{2} & \textbf{11\%} \\
Useful & \textbf{1} & \textbf{6\%} \\
\\
\textbf{What did you dislike about this card methodology?} & \textbf{\#}   & \textbf{\%} \\
\hline \hline
Many cards / rules to remember & \textbf{6} & \textbf{33\%}  \\
Programming difficulties / complexity & \textbf{4} & \textbf{22\%}  \\
Control flow (hand system of sequential vs. parallel) & \textbf{2} & \textbf{11\%} \\
Limited constructs (e.g., missing programming features) & \textbf{2} & \textbf{11\%}\\
Size of physical layout & \textbf{2} & \textbf{11\%} \\
Aspects of visual design & \textbf{1} & \textbf{6\%} \\
Difficulty locating physical cards & \textbf{1}& \textbf{6\%}  \\
Inconsistencies & \textbf{1} & \textbf{6\%} \\
Not intuitive & \textbf{1} & \textbf{6\%} \\
\\
\textbf{What improvements would you make to this system?} & \textbf{\#} & \textbf{\%}\\
\hline \hline
iPad app / computerize / GUI & \textbf{7} & \textbf{39\%} \\
Add cards (support new use cases, new functionality) & \textbf{3} & \textbf{17\%}\\
Replace cards (merge functionality into sophisticated cards) & \textbf{3} & \textbf{17\%}\\
Aspects of design / presentation & \textbf{2} & \textbf{11\%}  \\
Add programming constructs & \textbf{1} & \textbf{6\%}  \\
Improve consistency & \textbf{1} & \textbf{6\%} \\
Separate ``basic'' from ``advanced'' cards & \textbf{1} & \textbf{6\%} \\
\\
\textbf{Which cards most confused you?} & \textbf{\#}& \textbf{\%} \\
\hline \hline
Control flow (e.g., \textul{Branch} / \textul{Any} cards, \textul{Wait}) & \textbf{6}  & \textbf{33\%} \\
Movement (i.e., \emph{Movement} cards + related inputs / modifiers) &  \textbf{5}  & \textbf{28\%} \\
Non-movement actions (e.g, \emph{Tech} cards) & \textbf{3} & \textbf{17\%} \\
Computation (i.e., \emph{Think} cards) & \textbf{2} & \textbf{17\%} \\
\\
\textbf{Which cards were most useful to you?} & \textbf{\#}& \textbf{\%} \\
\hline \hline
Movement (i.e., \emph{Movement} cards + related inputs / modifiers) &  \textbf{13} & \textbf{72\%} \\
Computation (i.e., \emph{Think} cards) & \textbf{5}  & \textbf{28\%} \\
Control flow (e.g., \textul{Branch} / \textul{Any} cards, \textul{Wait}) & \textbf{5}  & \textbf{28\%} \\
Non-movement actions (e.g, \emph{Tech} cards) & \textbf{4} & \textbf{22\%} \\
\end{tabularx}
}
\caption{Post-study questionnaire free response after coding.}
\label{tbl:post_free}
\vspace{-1em}
\end{table}

\subsection{Expressiveness}
\label{sec:expressiveness}

Participants' ambitions in the ``create your own scenario'' task varied widely. Solutions ranged from simple one-hand programs with four cards to complex programs with many hands and 27 cards -- far more elaborate than any of our training or evaluation scenarios. Content ranged from practical (e.g., watering plants) to fantastical (e.g., killing zombies). In the post-study questionnaire, participants rated that they felt drone cards are expressible enough to implement ``a great deal'' (50\%) or ``some'' (50\%) drone scenarios (Table~\ref{tbl:post}).

\subsection{Engagement}

Overwhelmingly, participants felt ``very positive'' or ``generally positive'' (94\%) about the drone cards, and found the process ``very enjoyable'' or ``generally enjoyable'' (94\%). Only one participant was ``generally negative.'' (Table~\ref{tbl:post}). Two participants explicitly identified our approach as ``fun'' or ``playful'' in free response (Table~\ref{tbl:post_free}).

Participants did find several disagreeable aspects, especially the number of cards to learn and complexity of programming. Many offered suggestions for improvement; chiefly, digitizing the process through an app (39\%) or by making specific changes to the cards (e.g. adding or replacing cards).

\section{Discussion}

Our results generally confirm the efficacy of our card-based programming model. Our primary metrics of ease, expressiveness, and engagement rated favorably for most participants. Some participants were extremely positive, even after completing a demanding two-hour test session.
Though, perceptions were not uniform. A few participants found the approach unintuitive, complex, or inferior to traditional programming. Likely, these impressions are unavoidable; although many people enjoy card games, others find them complex, frustrating, or simply not fun. Overall, we find substantial evidence that card-based programming provides a useful abstraction to control complex systems such as drones.

%\subsubsection{Advantages and Disadvantages of the Card Interface}
%
%Our approach blends elements from visual, event-driven, and constraint-based programming paradigms and the structure of collectible card games. Despite reuse of known aspects, we believe our integration is novel and yields a powerful primitive that allows potential drone programmers to be able to create and manipulate drone programs easily. Our card designs and game structure allow complex programs to be built and executed. While we concede some some low-level programming constructs (namely, branching and logic operators), we focus the design around higher level actions and goals. In addition, reusable functions can be developed using the deck abstraction -- a drone programmer may develop a set of decks to perform different kinds of tasks, and then sequence those decks in a single drone program.

\subsubsection{Applicability to Other IoT Devices}

The card-based programming metaphor can be applied to domains beyond drones, and we readily see applicability to robots and other IoT devices that possess a capacity for sensing, thinking, acting, and communicating. The abstraction of an \emph{Action} card for performing behaviors, coupled with \emph{Token} cards as a serializing API to the underlying hardware, cover the capabilities of action and communication. The data flow provided by \emph{Input} cards and \emph{yields} can be used for sensing capabilities, and the decoupling of the \emph{use} of a \emph{Think} card from its underlying implementation shift the burden of managing difficult computation from the end-user to the implementer of the hardware tokens.

As a concrete example, TJBot~\cite{Dibia:2017} is an open-source paper robot designed to teach people about building cognitive systems. The computation involved in "bringing TJBot to life" involves the use of IBM Watson services to make him speak, listen, converse, and understand language and emotion. Cards for each of these facilities could be developed, abstracting the technical details of how (using machine learning to generate speech) and where (in the cloud) this behavior is performed from the end user who simply wishes to make their robot speak. The onus of responsibility for implementing these technical details (accessing Watson APIs via an SDK) lies with the developer in the implementation of TJBot's token cards.

\subsubsection{Real-world Considerations}

Up to this point, our presentation of our card-based programming metaphor has been purely theoretical -- the design and evaluation of the cards was conducted on paper. In order to understand whether the system would actually work for programming a real drone, we have implemented our system as a series of Swift frameworks. These frameworks are responsible for the construction and execution of card-based programs, as well as the specific behaviors of our drone cards and their interface to a real drone API (DJI). Overall we have found the implementation to be a straightforward task, but we learned a few key lessons.

\textbf{Separation of functionality.} A clean separation of functionality makes the card metaphor scale across different devices. Our implementation consists of these separate frameworks, with each subsequent framework dependent on the ones listed prior:
\begin{itemize}
\item \emph{CardKit}: specification of card programs (e.g. the ``compiler'').
\item \emph{CardKit Runtime}: execution engine for card programs, manages threaded execution of \emph{Action} cards, yield dataflow, hand branching logic, and error handling.
\item \emph{Drone Cards}: definition of drone-specific \emph{Action}, \emph{Input}, and \emph{Token} cards (e.g. \textul{FlyTo}, \textul{TakePhoto}), plus implementation of \emph{Action} cards against generic token interfaces.
\item \emph{Drone Tokens}: implementation of drone token cards against a specific drone API (e.g. the hardware interface layer). Support for new drone APIs is achieved by creating \emph{only} additional drone token implementations.
\end{itemize}

\emph{Action} card implementations communicate with \emph{Token} implementations using a well-defined, generic interface; thus, our set of drone cards can be used across multiple drone APIs or simulators by having separate implementations of just the token cards. Or, a wholly new kind of device can be supported (e.g. TJBot~\cite{Dibia:2017}) by defining a new set of cards and token implementations, without having to change the underlying specification and execution engines.

\textbf{Error handling.} Errors happen in code. To cope with this, we implemented an emergency stop feature~\cite{Steinfeld:2004} that can be triggered either by \emph{Action} or \emph{Token} cards. Emergency stop halts execution of the card program and signals token implementations to recover from the error. For example, the e-stop handler of the \textul{Movement} token triggers a landing.

\textbf{Platform mobility.} During development of the frameworks, we recognized that the task of \emph{constructing} a card program may occur at a different time or on a different device than the task of \emph{executing} it. Thus, not only did we map these two tasks into different frameworks, we also strongly enforced that the data structures comprising a card program be \emph{serializable} and \emph{portable}. In our frameworks, card programs are serializable to JSON, enabling them to be created on one device (e.g. a laptop) and executed on another (e.g. an iPad).

We have open sourced our frameworks~\footnote{https://github.com/CardKit} and encourage developers to build end-user apps that allow people to explore and create their own card-based programs for drones or other IoT devices.

\subsection{Limitations}

Although the spirit of the card programming metaphor is to be accessible to non-programmers, our evaluation used participants who were highly skilled in computational thinking. Given our desire to evaluate the expressiveness and effectiveness of our system, controlling the technical background of participants enabled us to understand whether difficulties in solving the programming tasks were due to failures of our own system rather than failures of participants to grasp the mechanics of the task. We are eager for future evaluations of our system with new populations of non-programmers and non-technical users to fully understand whether it fully lives up to its intended design.

Additionally, our system is not meant to enable the development of \emph{arbitrary} programs. We do not aim to provide a Turing-complete programming language; rather, we are satisfied if our model is able to express some percentage of the possible drone scenarios easily. The ``Create Your Own Scenario'' task aimed to test the limits of peoples' creativity to find breakdowns or limitations in our model, and participants came up with many scenarios we never anticipated. Participants also rated our model as having a high degree of expressiveness, giving us confidence in its utility.

%\subsubsection{The Physical Printed Medium}
%
%We decided to focus our research on using printed cards rather than digital cards because we were able to rapidly iterate on the game structure and card design through the physical medium. In addition, we found that physical demands of the card interface were low relative to the cognitive demands, confirmed by the TLX sub-scale scores (Figure~\ref{fig:tlx_all}(b)). Our ultimate vision is of a digital implementation, as this would facilitate the construction of programs (e.g. by making it easier to search for specific cards), and it would enable actual compilation and execution of drone card programs.
%
%Even when a digitized representation is available, we imagine that some users might prefer to work with paper cards. Several design choices support this use case: (1) the manner we recommend for laying out hands, (2) the location of card class and name; and (3) color coding. We anticipate an app able to take a photo of an entire card layout (all hands), applying OCR to recognize card names, and then compiling and uploading the program to the drone. As a final step, specific settings for \textul{Input} and \textul{Modifier} cards would be filled in by the drone operator (e.g., specifying a speed for \textul{Modifier:Speed}).

\section{Conclusion}

Drones are just beginning to revolutionize industries and day-to-day life. However, their use today is generally constrained to a fraction of their overall potential. We propose a card-based programming paradigm to automate diverse use cases in an intuitive, expressive, and fun manner. Heartened by participants' generally positive response to our paper prototype, we have developed and released CardKit, an open-source framework for card-based programming.

%\newpage

% BALANCE COLUMNS
\balance{}

% REFERENCES FORMAT
% References must be the same font size as other body text.
\bibliographystyle{SIGCHI-Reference-Format}
\bibliography{cardkit_ibm}

%%% -*-BibTeX-*-
%%% Do NOT edit. File created by BibTeX with style
%%% ACM-Reference-Format-Journals [18-Jan-2012].

\begin{thebibliography}{00}

%%% ====================================================================
%%% NOTE TO THE USER: you can override these defaults by providing
%%% customized versions of any of these macros before the \bibliography
%%% command.  Each of them MUST provide its own final punctuation,
%%% except for \shownote{}, \showDOI{}, and \showURL{}.  The latter two
%%% do not use final punctuation, in order to avoid confusing it with
%%% the Web address.
%%%
%%% To suppress output of a particular field, define its macro to expand
%%% to an empty string, or better, \unskip, like this:
%%%
%%% \newcommand{\showDOI}[1]{\unskip}   % LaTeX syntax
%%%
%%% \def \showDOI #1{\unskip}           % plain TeX syntax
%%%
%%% ====================================================================

\ifx \showCODEN    \undefined \def \showCODEN     #1{\unskip}     \fi
\ifx \showDOI      \undefined \def \showDOI       #1{{\tt DOI:}\penalty0{#1}\ }
  \fi
\ifx \showISBNx    \undefined \def \showISBNx     #1{\unskip}     \fi
\ifx \showISBNxiii \undefined \def \showISBNxiii  #1{\unskip}     \fi
\ifx \showISSN     \undefined \def \showISSN      #1{\unskip}     \fi
\ifx \showLCCN     \undefined \def \showLCCN      #1{\unskip}     \fi
\ifx \shownote     \undefined \def \shownote      #1{#1}          \fi
\ifx \showarticletitle \undefined \def \showarticletitle #1{#1}   \fi
\ifx \showURL      \undefined \def \showURL       #1{#1}          \fi

\bibitem{Bau:2017}
{David Bau}, {Jeff Gray}, {Caitlin Kelleher}, {Josh Sheldon}, {and} {Franklyn
  Turbak}. 2017.
\newblock \showarticletitle{Learnable Programming: Blocks and Beyond}.
\newblock {\em Commun. ACM\/} {60}, 6 (May 2017), 72--80.
\newblock
\showISSN{0001-0782}
\showDOI{%
\url{http://dx.doi.org/10.1145/3015455}}


\bibitem{brooke1996sus}
{John Brooke}. 1996.
\newblock \showarticletitle{SUS-A quick and dirty usability scale}.
\newblock {\em Usability evaluation in industry\/} {189}, 194 (1996), 4--7.
\newblock


\bibitem{Carlisle:2005}
{Martin~C. Carlisle}, {Terry~A. Wilson}, {Jeffrey~W. Humphries}, {and}
  {Steven~M. Hadfield}. 2005.
\newblock \showarticletitle{RAPTOR: A Visual Programming Environment for
  Teaching Algorithmic Problem Solving}. In {\em Proceedings of the 36th SIGCSE
  Technical Symposium on Computer Science Education} {\em (SIGCSE '05)}. ACM,
  New York, NY, USA, 176--180.
\newblock
\showISBNx{1-58113-997-7}
\showDOI{%
\url{http://dx.doi.org/10.1145/1047344.1047411}}


\bibitem{Dibia:2017}
{Victor~C. Dibia}, {Maryam Ashoori}, {Aaron Cox}, {and} {Justin~D. Weisz}.
  2017.
\newblock \showarticletitle{TJBot: An Open Source DIY Cardboard Robot for
  Programming Cognitive Systems}. In {\em Proceedings of the 2016 CHI
  Conference Extended Abstracts on Human Factors in Computing Systems} {\em
  (CHI EA '17)}. ACM, New York, NY, USA, 381--384.
\newblock
\showISBNx{978-1-4503-4656-6}
\showDOI{%
\url{http://dx.doi.org/10.1145/3027063.3052965}}


\bibitem{DJI:2017}
{DJI}. 2017 (accessed September 5, 2017).
\newblock {\em DJI Developer}.
\newblock
\newblock
\shownote{\url{http://developer.dji.com}.}


\bibitem{NodeRED:2017}
{JS Foundation}. 2017 (accessed September 5, 2017).
\newblock {\em Node-RED}.
\newblock
\newblock
\shownote{\url{http://nodered.org}.}


\bibitem{hart1988development}
{Sandra~G Hart} {and} {Lowell~E Staveland}. 1988.
\newblock \showarticletitle{Development of NASA-TLX (Task Load Index): Results
  of empirical and theoretical research}.
\newblock {\em Advances in psychology\/}  {52} (1988), 139--183.
\newblock


\bibitem{Tickle:2017}
{Tickle~Labs Inc.} 2017 (accessed September 5, 2017).
\newblock {\em Tickle: Program Star Wars BB-8, Drones, Arduino, LEGO, Dash \&
  Dot, Sphero, Robots, Hue, Scratch, Swift, and Smart Homes on your iPhone and
  iPad}.
\newblock
\newblock
\shownote{\url{http://tickleapp.com}.}


\bibitem{Ingalls:1988}
{Dan Ingalls}, {Scott Wallace}, {Yu-Ying Chow}, {Frank Ludolph}, {and} {Ken
  Doyle}. 1988.
\newblock \showarticletitle{Fabrik: A Visual Programming Environment}. In {\em
  Conference Proceedings on Object-oriented Programming Systems, Languages and
  Applications} {\em (OOPSLA '88)}. ACM, New York, NY, USA, 176--190.
\newblock
\showISBNx{0-89791-284-5}
\showDOI{%
\url{http://dx.doi.org/10.1145/62083.62100}}


\bibitem{Malan:2007}
{David~J. Malan} {and} {Henry~H. Leitner}. 2007.
\newblock \showarticletitle{Scratch for Budding Computer Scientists}. In {\em
  Proceedings of the 38th SIGCSE Technical Symposium on Computer Science
  Education} {\em (SIGCSE '07)}. ACM, New York, NY, USA, 223--227.
\newblock
\showISBNx{1-59593-361-1}
\showDOI{%
\url{http://dx.doi.org/10.1145/1227310.1227388}}


\bibitem{maloney2004scratch}
{John Maloney}, {Leo Burd}, {Yasmin Kafai}, {Natalie Rusk}, {Brian Silverman},
  {and} {Mitchel Resnick}. 2004.
\newblock \showarticletitle{Scratch: a sneak preview [education]}. In {\em
  Creating, Connecting and Collaborating through Computing, 2004. Proceedings.
  Second International Conference on}. IEEE, 104--109.
\newblock


\bibitem{Myers:1986}
{B.~A. Myers}. 1986.
\newblock \showarticletitle{Visual Programming, Programming by Example, and
  Program Visualization: A Taxonomy}. In {\em Proceedings of the SIGCHI
  Conference on Human Factors in Computing Systems} {\em (CHI '86)}. ACM, New
  York, NY, USA, 59--66.
\newblock
\showISBNx{0-89791-180-6}
\showDOI{%
\url{http://dx.doi.org/10.1145/22627.22349}}


\bibitem{Parrot:2017}
{Parrot}. 2017 (accessed September 5, 2017).
\newblock {\em Parrot for Developers}.
\newblock
\newblock
\shownote{\url{http://developer.parrot.com}.}


\bibitem{siegel2003sense}
{Mel Siegel}. 2003.
\newblock \showarticletitle{The sense-think-act paradigm revisited}. In {\em
  Robotic Sensing, 2003. ROSE'03. 1st International Workshop on}. IEEE, 5--pp.
\newblock


\bibitem{sousa2012dataflow}
{Tiago~Boldt Sousa}. 2012.
\newblock \showarticletitle{Dataflow programming concept, languages and
  applications}. In {\em Doctoral Symposium on Informatics Engineering}, Vol.
  130.
\newblock


\bibitem{Steinfeld:2004}
{A. Steinfeld}. 2004.
\newblock \showarticletitle{Interface lessons for fully and semi-autonomous
  mobile robots}. In {\em Robotics and Automation, 2004. Proceedings. ICRA '04.
  2004 IEEE International Conference on}, Vol.~3. 2752--2757 Vol.3.
\newblock
\showISSN{1050-4729}
\showDOI{%
\url{http://dx.doi.org/10.1109/ROBOT.2004.1307477}}


\bibitem{Ur:2014}
{Blase Ur}, {Elyse McManus}, {Melwyn Pak Yong~Ho}, {and} {Michael~L. Littman}.
  2014.
\newblock \showarticletitle{Practical Trigger-action Programming in the Smart
  Home}. In {\em Proceedings of the SIGCHI Conference on Human Factors in
  Computing Systems} {\em (CHI '14)}. ACM, New York, NY, USA, 803--812.
\newblock
\showISBNx{978-1-4503-2473-1}
\showDOI{%
\url{http://dx.doi.org/10.1145/2556288.2557420}}


\bibitem{whitley1997visual}
{Kirsten~N. Whitley}. 1997.
\newblock \showarticletitle{Visual programming languages and the empirical
  evidence for and against}.
\newblock {\em Journal of Visual Languages \& Computing\/} {8}, 1 (1997),
  109--142.
\newblock


\bibitem{young1995cantata}
{Mark Young}, {Danielle Argiro}, {and} {Steven Kubica}. 1995.
\newblock \showarticletitle{Cantata: visual programming environment for the
  Khoros system}.
\newblock {\em ACM SIGGRAPH Computer Graphics\/} {29}, 2 (1995), 22--24.
\newblock


\bibitem{zhao2009}
{Shengdong Zhao}, {Koichi Nakamura}, {Kentaro Ishii}, {and} {Takeo Igarashi}.
  2009.
\newblock \showarticletitle{Magic Cards: A Paper Tag Interface for Implicit
  Robot Control}. In {\em Proceedings of the SIGCHI Conference on Human Factors
  in Computing Systems} {\em (CHI '09)}. ACM, New York, NY, USA, 173--182.
\newblock
\showISBNx{978-1-60558-246-7}
\showDOI{%
\url{http://dx.doi.org/10.1145/1518701.1518730}}


\end{thebibliography}

\end{document}